\documentclass[aps,pra,floatfix,showpacs,onecolumn,tightenlines]{revtex4}

\usepackage{papersstyle}

\newcommand{\Poincare}{Poincar\'e}

\newlength{\manualrowheight}
\newcommand{\manualrowspace}[1][1]%
  {\makebox[0mm]{\normalsize\setlength{\manualrowheight}{#1\baselineskip}
  \vphantom{\rule{0mm}{\manualrowheight}}}}

\newcommand{\mc}[1]{\ensuremath{\mathcal{#1}}}		
\newcommand{\undertilde}[1]{\rlap{\smash{\lower 2ex \hbox{$\tilde{\hphantom{#1}}$}}}#1}
\newcommand{\lundertilde}[1]{\rlap{\smash{\lower 2.2ex \hbox{$\tilde{\hphantom{#1}}$}}}#1}
\newcommand{\uvec}[1]{\ensuremath{\undertilde{#1}}}		

\newcommand{\bra}[1]{\langle#1|}		
\newcommand{\ket}[1]{|#1\rangle}		
\newcommand{\expect}[1]{\ensuremath{\langle#1\rangle}}  
\newcommand{\conj}{^{\ast}}		
\newcommand{\br}[1]{\left(#1\right)}		
\newcommand{\sqbr}[1]{\left[#1\right]}		
\newcommand{\cbr}[1]{\left\{#1\right\}}		
\newcommand{\Tr}[1]{\mathrm{Tr}\cbr{#1}}
\newcommand{\textTr}[1]{\mathrm{Tr}\{#1\}}		
\newcommand{\nn}{\nonumber}

\begin{document}

\title{Errors in quantum tomography: diagnosing systematic versus statistical errors}

\author{Nathan~K.~Langford}
\affiliation{Department of Physics, Royal Holloway University of London, Egham, Surrey, TW20 0EX, United Kingdom\\
nathan.langford@rhul.ac.uk}

\date{\today}

\begin{abstract}
A prime goal of quantum tomography is to provide quantitatively rigorous characterisation of quantum systems, be they states, processes or measurements, particularly for the purposes of trouble-shooting and benchmarking experiments in quantum information science.  A range of techniques exist to enable the calculation of errors, such as Monte-Carlo simulations, but their quantitative value is arguably fundamentally flawed without an equally rigorous way of authenticating the quality of a reconstruction to ensure it provides a reasonable representation of the data, given the known noise sources.  A key motivation for developing such a tool is to enable experimentalists to rigorously diagnose the presence of technical noise in their tomographic data.  In this work, I explore the performance of the chi-squared goodness-of-fit test statistic as a measure of reconstruction quality.  I show that its behaviour deviates noticeably from expectations for states lying near the boundaries of physical state space, severely undermining its usefulness as a quantitative tool precisely in the region which is of most interest in quantum information processing tasks.  I suggest a simple, heuristic approach to compensate for these effects and present numerical simulations showing that this approach provides substantially improved performance.
\end{abstract}


\maketitle

One of the greatest challenges associated with trying to demonstrate a quantum information processing (QIP) protocol experimentally is to be able to verify and quantify its successful operation.  In some cases, such as Bell~\cite{ClauserJF1969} and steering~\cite{WisemanHM2007} tests of nonlocal quantum phenomena and Kochen-Specker tests of noncontextuality~\cite{KlyachkoAA2008}, this can be achieved by violating a measurement inequality.  In other cases, such as Shor's factoring algorithm~\cite{ShorPW1994}, the success of the protocol can be readily tested after the fact via a simple test of the ``correctness'' of the outcome or answer.  Often the answer is not so clear cut~\cite{WhiteAG2007}, however, and quantum tomography can be a valuable tool for achieving this goal.  This might, for example, take the form of measuring the process directly via quantum process tomography (QPT)~\cite{ChuangIL1997}, or characterising the state at key stages of the protocol via quantum state tomography (QST)~\cite{SmitheyDT1993}.

At its heart, tomography aims to be a quantitative tool for providing information about the system being studied.  When wishing to use it to quantify the success or performance of a real-world quantum protocol in a precise way, it is therefore critical to have a rigorous and comprehensive recipe for dealing with the effects of noise.  The first and most obvious ingredient of such a recipe is the necessity to have a way to calculate errors either in the tomographic reconstruction itself, or in subsequently derived physical parameters.  In Bayesian approaches to tomography, such as Bayesian mean estimation~\cite{Blume-KohoutR2010a} or Kalman filtering reconstruction~\cite{AudenaertKMR2009}, the method provides errors automatically as part of the reconstruction output.  Currently, the most commonly used method of tomographic reconstruction, however, is maximum-likelihood estimation~\cite{HradilZ1997, JamesDFV2001}.  The formal output of this approach is a single quantum state or process, the one which mathematically maximises the likelihood function, with no intrinsic uncertainty.  Maximum-likelihood tomography therefore needs to be augmented in some way to allow the experimenter to estimate errors.  Perhaps the most common way to do this is to use Monte-Carlo simulations, although several other methods have also been proposed for doing this in recent years (e.g.,~\cite{Blume-KohoutR2010b, ChristandlM2012, Blume-KohoutR2012}).  One feature typical to most methods for determining errors, is that it is first necessary to make assumptions about the precise error model or form of noise affecting the individual tomographic measurements.  But how do we know when this assumption is actually valid?

This question highlights the fact that calculating error bars is only half the story.  Like any data fitting technique, a critical component of tomography must be to have a means of establishing whether the fit or model is in any way a meaningful quantitative representation of the original data.  This aspect has been largely ignored until recently~\cite{LangfordNK2007phd, MoroderT2012}, but is particularly important in this context.  In tomography, the measurements are sufficiently complex and conceptually removed from the more abstract theoretical entities which we use to describe the underlying system, like the state and process matrices, that looking directly at the raw data often provides virtually no information to the experimentalist.  The tomographic reconstruction itself is often the most meaningful graphical representation and serves as a proxy for the data when interpreting results.  Furthermore, when trying to characterise a quantum system with a view to determining fault tolerance, the margins are usually so tight that approximate measures are simply not good enough.  It is therefore critical to know whether the reconstruction is reasonable and how much confidence can be placed in any results derived from it.

Perhaps the most important motivation for being able to assess the quality of a tomographic reconstruction is that, in any real-world context, measurements are affected by multiple sources of error, both fundamental and technical in nature.  Fundamental noise arises from the fact that quantum mechanical measurements are inherently probabilistic: this leads to statistical errors for any measurement with a finite number of system copies.  But in practice, any experiment will also be affected by technical noise sources, errors that arise from inaccuracies or instabilities in the measurement apparatus.  To date, most errors reported in QIP experiments consider only fundamental noise sources and ignore errors which arise from technical noise.  This is reasonably well justified in cases where statistical counting errors dominate, such as multiphoton polarisation tomography, which often involves low count rates and high-precision measurements.   However, it is likely to be much less reasonable in contexts such as superconducting qubit experiments, where systematic errors and imprecision in measurement settings can strongly dominate over statistical errors, making it a critical open problem to develop methods for incorporating both fundamental and technical noise sources into tomographic error calculations.  In either case, however, if tomography is to provide meaningful and rigorous analysis, it is necessary to justify its assumptions quantitatively.

A range of statistical tools exists for addressing such questions (see, e.g.,~\cite{Mood, Taylor}), under the broader umbrella of hypothesis testing, such as likelihood-ratio tests~\cite{Blume-KohoutR2010c, MoroderT2012} and chi-squared tests.  In this paper, I explore the standard chi-squared goodness-of-fit test statistic as a measure of tomographic reconstruction quality and examine its performance from an operational perspective.  I show that, while the statistic performs well for full-rank mixed states, its behaviour becomes much more complex for states lying near the boundary of the space of physical states: in this region, making such measures quantitatively rigorous requires careful treatment of the constraints imposed by the tomographic reconstruction process.  In particular, I show that na\"ively assuming that the optimal matrix involves $d^2$ degrees of freedom leads to substantial over-diagnosis of technical noise, severely undermining the quantitative value of this and other similar measures.  Determining the precise effect of the physicality constraints on goodness-of-fit and hypothesis testing is extremely complex, because of the nature of the convex inequality constraints imposed by positivity.  In this work, however, I introduce a simple approach to addressing this problem, based on counting the number of free parameters in rank-deficient mixed states.  Applying this to the context of chi-squared tests, I define two related ways of analysing the chi-squared statistic to assess reconstruction quality and demonstrate that they perform substantially better in typical experimental situations than the na\"ive approach which simply ignores the effects of physicality constraints.  Finally, I discuss the usefulness and reliability of the suggested techniques in real-world applications.  In the context of photonic tomography carried out in the presence of fluctuations in source brightness, I show that they are able to diagnose relatively low-level fluctuations, even when they are smaller than the statistical counting noise.  I also briefly discuss how the chi-squared test statistic provides qualitative evidence for the value of using over-complete rather than minimally complete measurement sets.

\subsection{Maximum likelihood quantum state tomography---the basic framework}

In this section, I provide a brief description of a specific simple form of state tomography loosely following the treatment of QPT described in Ref.~\cite{NielsenChuang}.  Since there are straightforward correspondences between the different types of tomography (e.g., state and process), I will focus here on state tomography.

Consider the following scenario:  we have a source which produces identical copies of a quantum system in the state $\rho$ (dimension $d$) and we implement a particular quantum state tomography which is defined by a set of informationally complete measurement projectors $P_j$ in order to estimate the ``true state'' $\rho$.  Performing this tomography will yield a vector of measurement values $\uvec{\mc{N}}\equiv(\mc{N}_j)$ which can be used to reconstruct an estimated state, $\varrho$.  Throughout this paper, I use calligraphic letters and symbols to indicated measured or estimated values, and the notation that $\uvec{a}=(a_j)$ is a vector and $\tilde{a}=(a_{jk})$ is a matrix.  Since the true state can be written as a linear combination,
\begin{align}
\label{eq-orthonormal-basis}
\rho = \sum_{j=1}^{d^2} o_j O_j,
\end{align}
of some orthonormal basis of $d {\times} d$ matrices, $\{O_j\}$, the problem of estimating the quantum state reduces to the problem of estimating the $d^2$ coefficients $\uvec{o} \equiv (o_j)$.

The simplest way to estimate the underlying quantum state is to implement an algebraic linear inversion, which can be summarised as follows.  We first note that the measurement projectors can also be expanded in terms of the orthonormal basis:
\begin{align}
P_j = \sum_{k=1}^{d^2} q_{jk} O_k,
\end{align}
where the coefficients $\tilde{q}=(q_{jk}) = (\textTr{O_k^\dag P_j})$ can be determined without reference to measurement results.  These projectors also define expected measurement probabilities in terms of the unknown ``true state'', $p_j \equiv \Tr{P_j \rho}$, which, via some simple algebra, can be related to the orthonormal coefficients:
\begin{align}
\label{eq-linear-tomo1}
p_j = \Tr{P_j \rho}
= \Tr{P_j^\dag \rho} 
= \sum_k q_{jk}\conj \Tr{O_k^\dag \rho}
= \sum_k q_{jk}\conj o_k.
\end{align}
We can also define a measured ``probability'' in terms of measured values and a potentially unknown normalisation parameter, $\mc{P}_j \equiv \mc{N}_j / \overline{\mc{N}}$.  Making the standard frequentist assumption that this measured frequency is the best estimate of the underlying true-state probability, we can then invert Eq.~(\ref{eq-linear-tomo1}) above to give estimates of the $o_j$ coefficients:
\begin{align}
\label{eq-linear-tomo2}
\uvec{\mc{O}} \equiv \sqbr{\tilde{q}\conj}^{-1} \uvec{\mc{P}}.
\end{align}
Technically, the matrix $\tilde{q}$ is invertible if and only if the set of measurement projectors forms a linearly independent or minimally informationally complete set.  If not, the estimated coefficients can instead be calculated using the Moore-Penrose pseudoinverse~\cite{PenroseR1955}.  In particular, if the measurement set is over-complete, this then gives the least-squares solution to Eq.~(\ref{eq-linear-tomo2}).  Finally, we reconstruct an estimated density matrix via a modified form of Eq.~(\ref{eq-orthonormal-basis}):
\begin{align}
\label{eq-linear-tomo3}
\varrho \equiv \sum_{j=1}^{d^2} \mc{O}_j O_j.
\end{align}
Note that, if the normalisation parameter, $\overline{\mc{N}}$, is unknown, as is commonly the case in photonic tomography, for example, it is straightforwardly calculated at this stage as part of the tomographic reconstruction as follows.  Firstly, from the definition of $\mc{P}_j$, note that $\overline{\mc{N}} \uvec{\mc{O}} \equiv \sqbr{\tilde{q}\conj}^{-1} \uvec{\mc{N}}$.  Then, defining a new, unnormalised reconstructed matrix, $\bar{\varrho}$, by applying the inversion matrix directly to the vector of measured counts, it is simple to show that:
\begin{align}
\bar{\varrho} \equiv \sum_{j=1}^{d^2} (\sqbr{\tilde{q}\conj}^{-1} \uvec{\mc{N}})_j O_j = \overline{\mc{N}} \varrho,
\end{align}
where the trace of $\bar{\varrho}$ is equal to the unknown normalisation parameter, $\overline{\mc{N}}$, and the normalised part is just the standard reconstructed density matrix, $\varrho$.

\begin{figure}
\begin{center}\begin{tabular}{ccccc}
\includegraphics[width=37.5mm]{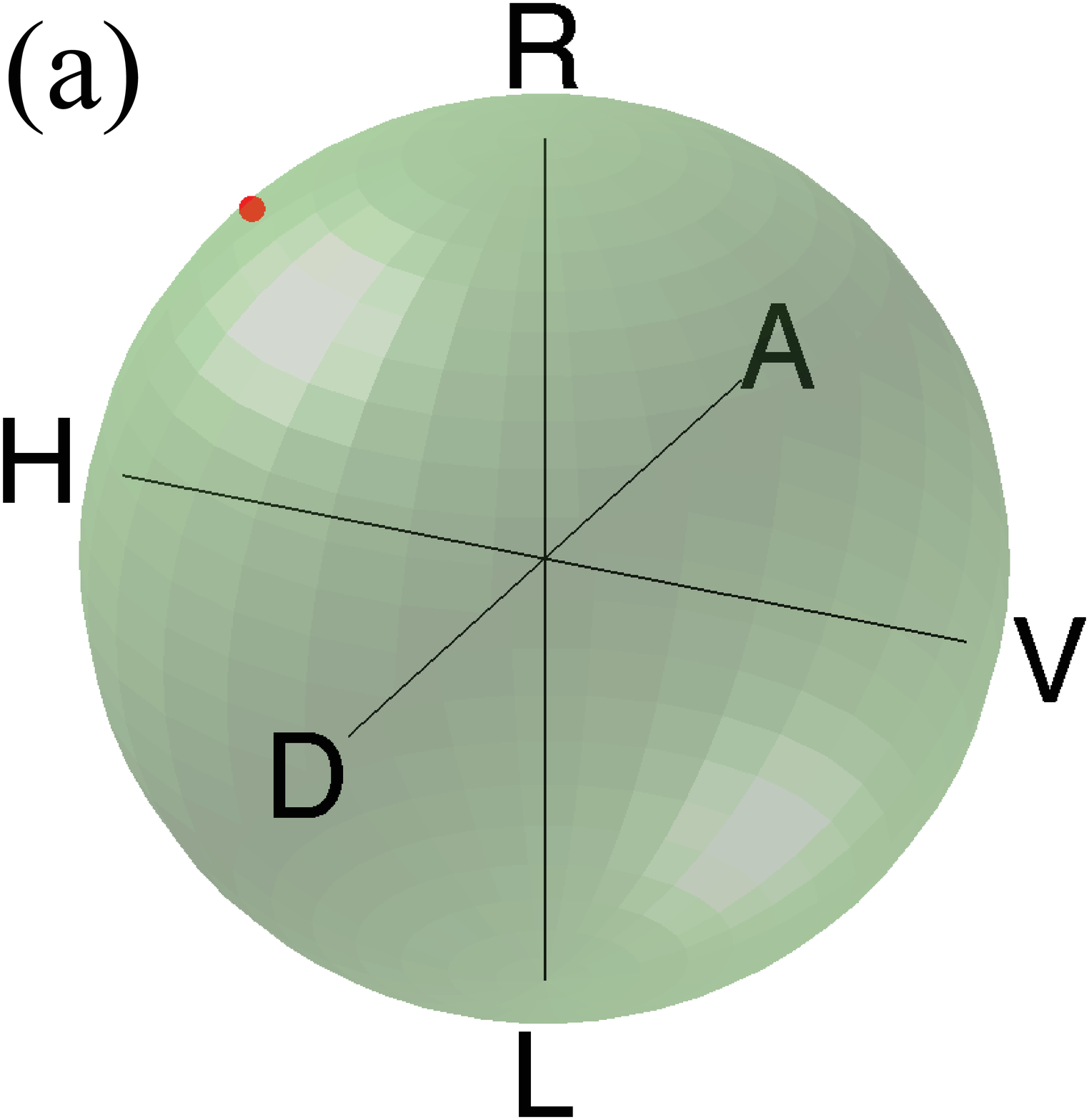} & \hspace{0.03 \columnwidth} &
\includegraphics[width=37.5mm]{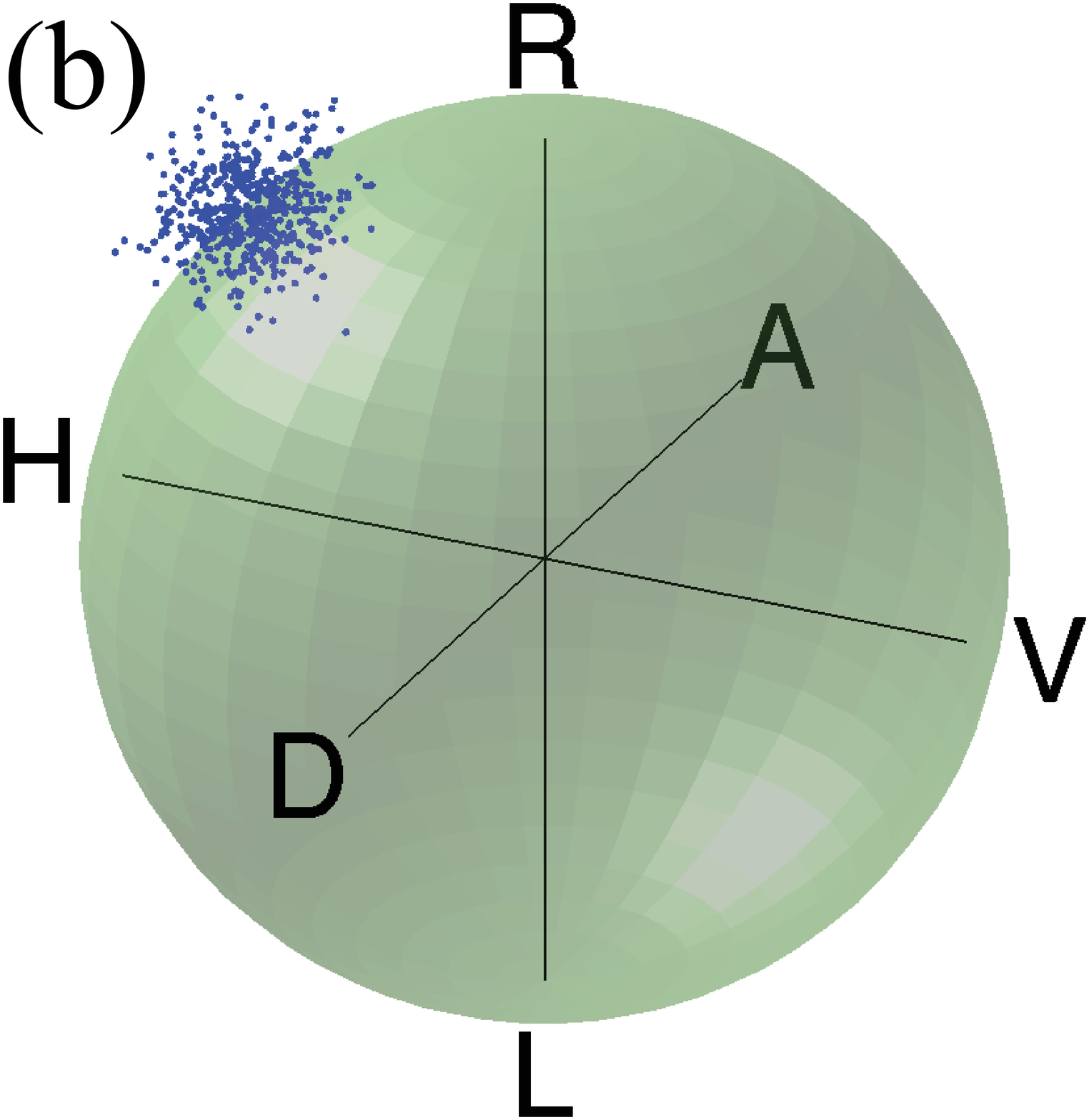} & \hspace{0.03 \columnwidth} &
\includegraphics[width=37.5mm]{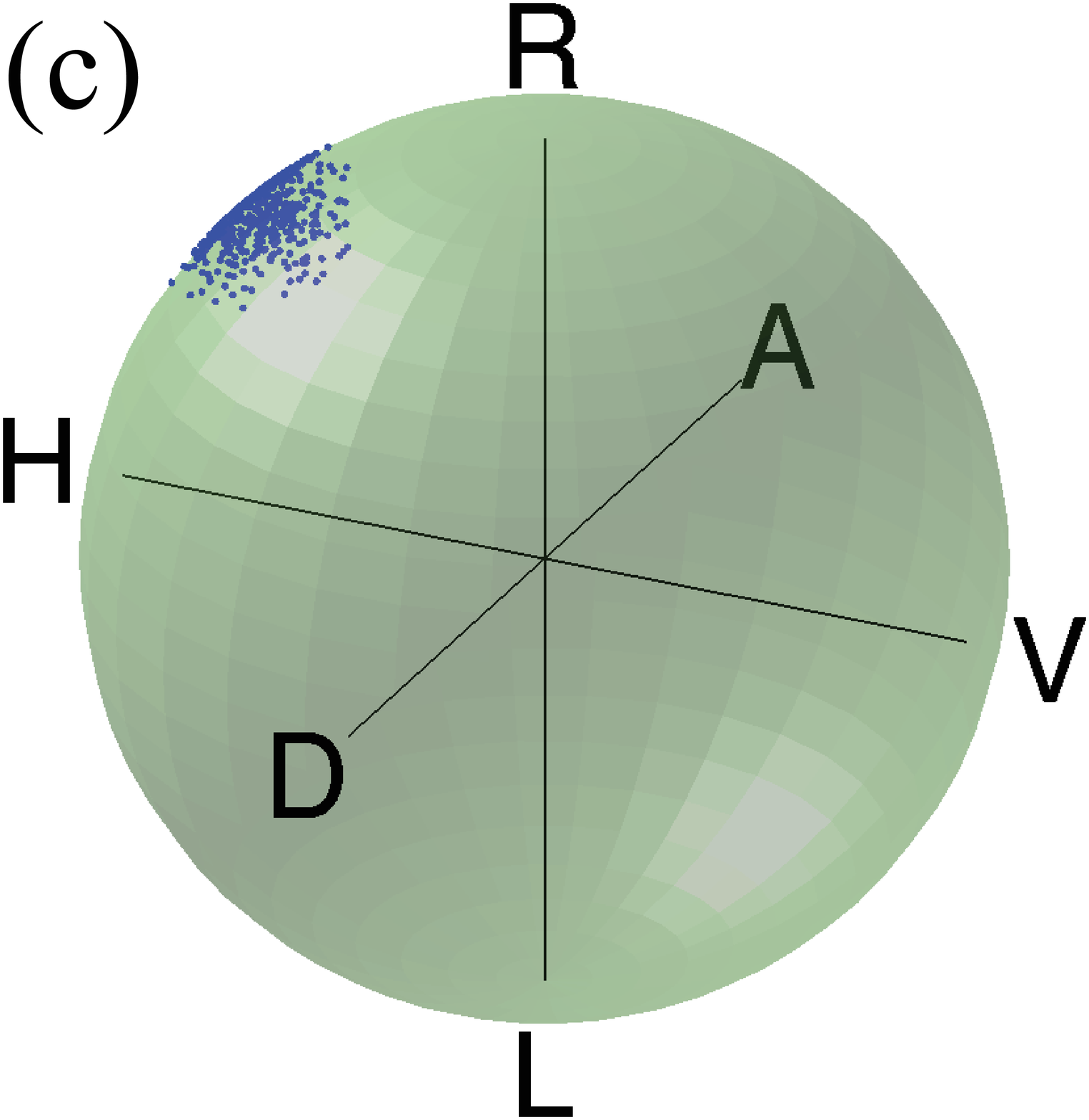}
\end{tabular}\end{center}
\caption{Simulations illustrating the effects of noise on linear inversion.  A known, nearly pure target state (a) is used to generate 500 sets of Poissonian-distributed noisy data with a mean photon flux of 100 photons per measurement setting, which are then reconstructed using a na\"ive linear inversion (b) and maximum likelihood estimation (c).  The linear inversion produces many non-physical states which lie outside the \Poincare\ sphere.}
\label{fig-linear-vs-maximum-likelihood}
\end{figure}

Equations~(\ref{eq-linear-tomo2}) and~(\ref{eq-linear-tomo3}) reveal the first problem that arises from noise in the measurements.  Because the measured values, $\mc{N}_j$, and therefore also the estimated coefficients, $\mc{O}_j$, necessarily contain noise, even if only as a result of the statistics of measurements made with a finite number of system copies, the linear inversion estimate from Eq.~(\ref{eq-linear-tomo3}) is not necessarily physical.  Specifically, while it is possible to ensure that it is Hermitian (by choosing all $O_j$ to be Hermitian), the reconstructed matrix may not be positive.  For a single qubit, this corresponds to the state estimate lying outside the Bloch (\Poincare) sphere [Fig.~\ref{fig-linear-vs-maximum-likelihood}].

The typical way to solve this problem is to use maximum likelihood tomography~\cite{HradilZ1997, JamesDFV2001}, an optimisation procedure which is very similar to common curve-fitting data analysis techniques.  Importantly, however, this requires no assumptions about the form of the measured state except that it be physical.  Instead, the goal is simply to search the entire space of allowed, physical density matrices for the one with the highest probability of generating the measurement results.  This probability is known as the likelihood function.  

In any fitting procedure, provided the measurements outnumber the fitting parameters, the optimal values of the fitting parameters will necessarily deviate slightly from the individual observed data points.  This is also true for maximum likelihood tomography.  In this case, however, imposing physicality constraints provides an additional cause for mismatch between the observed data and values predicted from the reconstructed state, even in the case of a minimally complete measurement set.

For a particular measurement record, $\uvec{\mc{N}}$, the likelihood function is~\cite{HradilZ1997}
\begin{align}
\mc{L}(\uvec{\mc{N}}|\rho) \propto \prod_j p_j^{\mc{N}_j} = \prod_j \Tr{P_j \rho}^{\mc{N}_j},
\end{align}
omitting prefactors that depend only on the particular measurement record and therefore do not influence the outcome of the optimisation over physical states.  Note that, rather than maximising the likelihood function, it is actually more common to carry out an equivalent minimisation of the negative log-likelihood penalty function:
\begin{align}
\label{eq-likelihood-multinomial}
\Pi(\uvec{\mc{N}},\rho) = -\log \mc{L}(\uvec{\mc{N}}|\rho) = -\sum_j \mc{N}_j \log (\Tr{P_j \rho}).
\end{align}
This type of optimisation falls in the special class of convex optimisation problems~\cite{BoydVandenberghe,KosutRL2004}.  While they still scale exponentially in the system size, such problems are nevertheless computationally efficient for a given system size and recent efforts combining convex optimisation with compressive sensing techniques have made substantial progress towards minimising both the experimental and numerical resources required for solving them~\cite{KosutRL2008,GrossD2010,FlammiaST2012,ShabaniA2011}.  Overall, the maximum-likelihood quantum state estimation optimisation problem can be summarised as:
\begin{align}
\nn \text{minimise} \quad& \Pi(\uvec{\mc{N}},\rho), \\
\text{subject to} \quad& \rho \ge 0 \;\; \& \;\; \Tr{\rho}=1,
\end{align}
where I have assumed that $\rho = \rho^\dag$ has been enforced by choosing all $O_j$ to be Hermitian.

This optimisation can now be implemented in different ways (e.g., Refs~\cite{HradilZ1997, JamesDFV2001,KosutRL2004}).  For this work, I now follow the method in Ref.~\cite{JamesDFV2001}, which assumes that the counts are sampled from Gaussian distributions with variances, $\sigma_j^2$, equal to the means of the distributions (as is true for the Poissonian statistics common in photonic tomography experiments).  In this case, the log-likelihood penalty function reduces to a weighted least squares form:
\begin{equation}
\label{eq-likelihood-gaussian}
\Pi(\uvec{\mc{N}},\rho,\overline{\mc{N}}) = \sum_j \frac{\br{\mc{N}_j - n_j(\rho,\overline{\mc{N}})}^2}{\sigma_j(\rho,\overline{\mc{N}})^2} = \sum_j \frac{\br{\mc{N}_j - n_j(\rho,\overline{\mc{N}})}^2}{n_j(\rho,\overline{\mc{N}})},
\end{equation}
where $n_j(\rho,\overline{\mc{N}}) = \overline{\mc{N}} \Tr{P_j \rho}$ and I have again explicitly included the normalisation parameter.  (Note that the more standard, linear weighted least squares problem would involve fixed variances which do not depend on the optimisation parameters.)  This can now be recast as follows as a particular form of convex optimisation called a semidefinite programme~\cite{BoydVandenberghe,DohertyAC2006,LangfordNK2007phd}.  Such problems are the focus of a large body of work and many good numerical routines are readily and freely available to solve them.

To do this, I follow the approach of Ref.~\cite{DohertyAC2006} (see Ref.~\cite{LangfordNK2007phd}), which has two key steps.  The first is to replace the nonlinear objective or penalty function with a linear objective function and a nonlinear constraint:
\begin{align}
\nn \text{minimise} \quad& y, \\
\nn \text{subject to} \quad& y - \sum_j \delta_j^2/n_j \ge 0, \\
& \rho \ge 0 \quad \& \quad \Tr{\rho}=1,
\end{align}
where $\delta_j = \delta(\uvec{\mc{N}},\rho,\overline{\mc{N}}) = \mc{N}_j - n_j(\rho,\overline{\mc{N}})$ are the \emph{residuals} between measured and expected counts and $y$ is a so-called \emph{slack variable}, which will achieve a minimum value for the same density matrix as the negative log-likelihood function.  Secondly, the nonlinear constraint is then replaced by an equivalent larger-dimensional linear matrix constraint to give the final optimisation problem:
\begin{align}
\nn \text{minimise} \quad& y, \\
\nn \text{subject to} \quad& \sqbr{ \begin{matrix}
y & \uvec{\delta}^T \\
\uvec{\delta} & \tilde{n}
\end{matrix} } \ge 0,
 \\
& \rho \ge 0 \quad \& \quad \Tr{\rho}=1,
\end{align}
where $\tilde{n}$ is the diagonal matrix defined according to $\br{\tilde{n}}_{jj} = n_j(\rho,\overline{\mc{N}})$.  This is now written explicitly in the form of a semidefinite programme.

Finally, if the normalisation parameter is unknown, it can again be calculated as part of the reconstruction process by simply removing the third constraint that $\Tr{\rho}=1$ and redefining $n_j(\rho,\overline{\mc{N}}) = \overline{\mc{N}} \Tr{P_j \rho} \equiv \Tr{P_j \bar{\rho}}$, where $\bar{\rho}$ is an unnormalised ``density matrix'' variable.  The optimisation will now output an unnormalised estimate, $\bar{\varrho}=\overline{\mc{N}} \varrho$, where the trace is again the unknown normalisation parameter and the normalised part is the expected maximum-likelihood estimate.

Throughout this paper, unless otherwise stated, I will focus on the context of photonic tomography, although the main concepts should generalise straightforwardly.  For concreteness, I will assume that the normalisation is an unknown parameter which is calculated as part of the reconstruction process, i.e., the output is the unnormalised density matrix, $\bar{\varrho}$.  I will also assume that data are collected using successive, single-projector measurements.  Therefore, considering only fundamental noise sources, for a given state, $\rho$, the measured counts $\mc{N}_j$ are Poissonian-distributed random variables with mean, $\expect{\mc{N}_j}=n_j(\bar{\rho})= \Tr{P_j \bar{\rho}}$, and variance, $\sigma_j^2(\bar{\rho})\equiv \expect{\br{\mc{N}_j - \expect{\mc{N}_j}}^2} {=}n_j$.

\subsection{Reconstruction quality: the chi-squared ``goodness of fit''}

In the absence of technical noise, given a set of measured data used to estimate a quantum state, we would expect the data to always agree with the predicted counts (as calculated from the reconstructed state) to within limits defined by fundamental statistical fluctuations.  We can therefore diagnose the presence of technical noise if we can show with some confidence that this is not the case.  This is exactly the scenario that is described by a standard chi-squared test.  We have a set of observed frequencies, $\{\mc{N}_j\}$, and a  set of expected outcome frequencies, $\{n_j(\bar{\varrho})\}$, with standard deviations, $\sigma_j^2(\bar{\varrho})$, and we can construct the standard chi-squared test statistic for measuring the ``goodness of fit'' as follows~\cite{Taylor,Mood}:
\begin{equation}
\label{eq-test-statistic}
X^2(\bar{\varrho}) = \sum_{j=1}^M \frac{ \br{\mc{N}_j - n_j(\bar{\varrho})}^2 }{ \sigma_j^2(\bar{\varrho}) } = \sum_{j=1}^M \frac{ \delta_j(\bar{\varrho})^2 }{ \sigma_j^2(\bar{\varrho}) }.
\end{equation}
Note that $X^2$ has the same form as Eq.~(\ref{eq-likelihood-gaussian}), the negative log-likelihood penalty function, $\Pi(\uvec{\mc{N}},\bar{\varrho})$, in the Gaussian approximation, which is exactly the limit in which chi-squared-based tests are valid.  Intuitively, this provides a motivation for the approach.  We would like to answer the question: ``\emph{How well} does the reconstructed density matrix fit the measured data?''  This is exactly what the likelihood function itself seeks to answer.  Other approaches, like likelihood-ratio tests, build more directly on this intuition.
                                                                                                                                                                                                                                                                                                                                                                                                                                                                                                   
Under appropriate conditions, the chi-squared test statistic, $X^2$, is distributed according to a $\chi^2$ distribution with $\kappa = M - c$ degrees of freedom (DOFs), where $c$ is the number of constraints imposed by the measurement scheme and the fitting of density-matrix parameters during the reconstruction process.  The mean and variance of the $\chi^2$ distribution are $\kappa$ and $2\kappa$, respectively~\cite{Mood}.  The same results can be obtained directly from Eq.~(\ref{eq-test-statistic}) if the residuals in the test statistic are Gaussian random variables.  For certain states, however, there may be individual measurements with very low counts, even if many system copies are measured overall.  For the case of Poissonian photon statistics, this introduces a slight correction to the variance, giving instead a value of $2\kappa + \sum_j 1/n_j$ \cite{Taylor, Note-PoissonianCorrection}.

In order to ensure validity of the chi-squared test, standard practice recommends that the expected counts for any given measurement should be large enough~\cite{Taylor, Mood}, normally greater than 5 or 10, or at least that that is true for the large majority of measurements.  If that condition is violated (which would be rare for the less stringent condition in practical situations), it is allowable to group data from different measurements into bins with sufficiently large total expected counts.  For systems where the measured counts arise from a well-defined number of total trials, such as ion-trapping experiments, the measurements would be described by multinomial statistics, making the Poissonian statistics used in the above definition only an approximation.  Here, as well as ensuring that there are enough expected counts in each bin, it is also important to ensure that no count represents too large a fraction of the results for a given setting.  However, since one count in each measurement setting in this context is always dependent on the others via the constraint on the total number of trials, it is always possible to choose to eliminate the large count (there can be only one) to ensure the Poissonian approximation remains valid~\cite{ThomasPeterN2011}.

Both mean and standard deviation values depend on the measurements and constraints for the system and therefore vary with system size and measurement recipe.  For everyday use, it is therefore generally more intuitive to use the reduced chi-squared test statistic, which I will often call simply the reconstruction \emph{quality} for compactness:
\begin{equation}
\label{eq-reduced-test-statistic}
Q(\bar{\varrho}) \equiv X^2_{\rm r}(\bar{\varrho}) = \frac 1\kappa X^2(\bar{\varrho}),
\end{equation}
which physically describes the mean-square residual count error (per degree of freedom) in the final reconstruction.  Assuming, as before, that the residuals are Gaussian random variables, the mean and variance of $Q$ are $1$ and $2/ \kappa$, respectively.

For larger values of $\kappa$, the quality parameter should generally be very close to 1 if the experimental noise is dominated by the known, fundamental statistical errors.  If the value is instead a sufficient amount larger than expected, it indicates that the statistical noise is not sufficient to explain the noise observed in the raw data.  More concretely, for a given reconstruction, one can calculate either the standard or reduced chi-squared test statistic and compare it with the expected value, or rather, with a cut-off value which determines a particular confidence level.  If the measured value is above the cut-off, then this diagnoses the presence of unexpected noise at that given confidence level.  While it is not necessarily obvious how to incorporate this information into the noise model used to make improved error estimates, it does provide a quantitative assessment of the fitting process itself and the quality of the measured data.

Up to this point, the discussion has progressed using basic statistical tools, but when applying these techniques to quantum tomography, the real difficulty that arises is how to deal with the effects of physicality constraints, more specifically, the inequality constraints imposed by requiring the reconstruction to be positive.  As positivity constraints come into play, the general effect is to confine the reconstruction onto a surface of reduced dimensionality at the boundary of physical state space, which tends to reduce the number of free parameters in the optimisation.  For example, for a single qubit, positivity forces points outside the Bloch sphere onto the surface of the sphere, a two-dimensional surface within a three-dimensional space [Fig.~\ref{fig-linear-vs-maximum-likelihood}].  But the number of constraints determines the entire shape of the test statistic's expected distribution via $\kappa$, the DOFs.  For example, while an arbitrary $d$-dimensional density matrix (unnormalised) contains $d^2$ independent parameters ($d^2{-}1$, if normalised), an arbitrary $d$-dimensional pure state can be specified by $2d-1$ real parameters (again, unnormalised).  Since the pure states lie at the boundary of the physical state space, this is where these effects are most likely to manifest.  Unfortunately, in quantum information processing applications, the goal is often to produce states which are as close as possible to pure states (or unitary operations in the case of quantum processes).  Such experiments are therefore almost always aiming to probe the region where these effects will be seen, especially as they become ever more reliable and the need for precise quantitative analysis becomes ever more critical.  Furthermore, in cases where measurement noise is large enough for physicality constraints to play a role, it is known that maximum likelihood tomography has a tendency to produce states that contain one or more zero eigenvalues, e.g., in one-qubit experiments, to produce pure states.  In the remainder of this paper, I will study this issue and how to resolve it.

\subsection{Performance of the $X^2$ test statistic}

\begin{figure}
\begin{center}\begin{tabular}{ccccc}
\includegraphics[width=45mm]{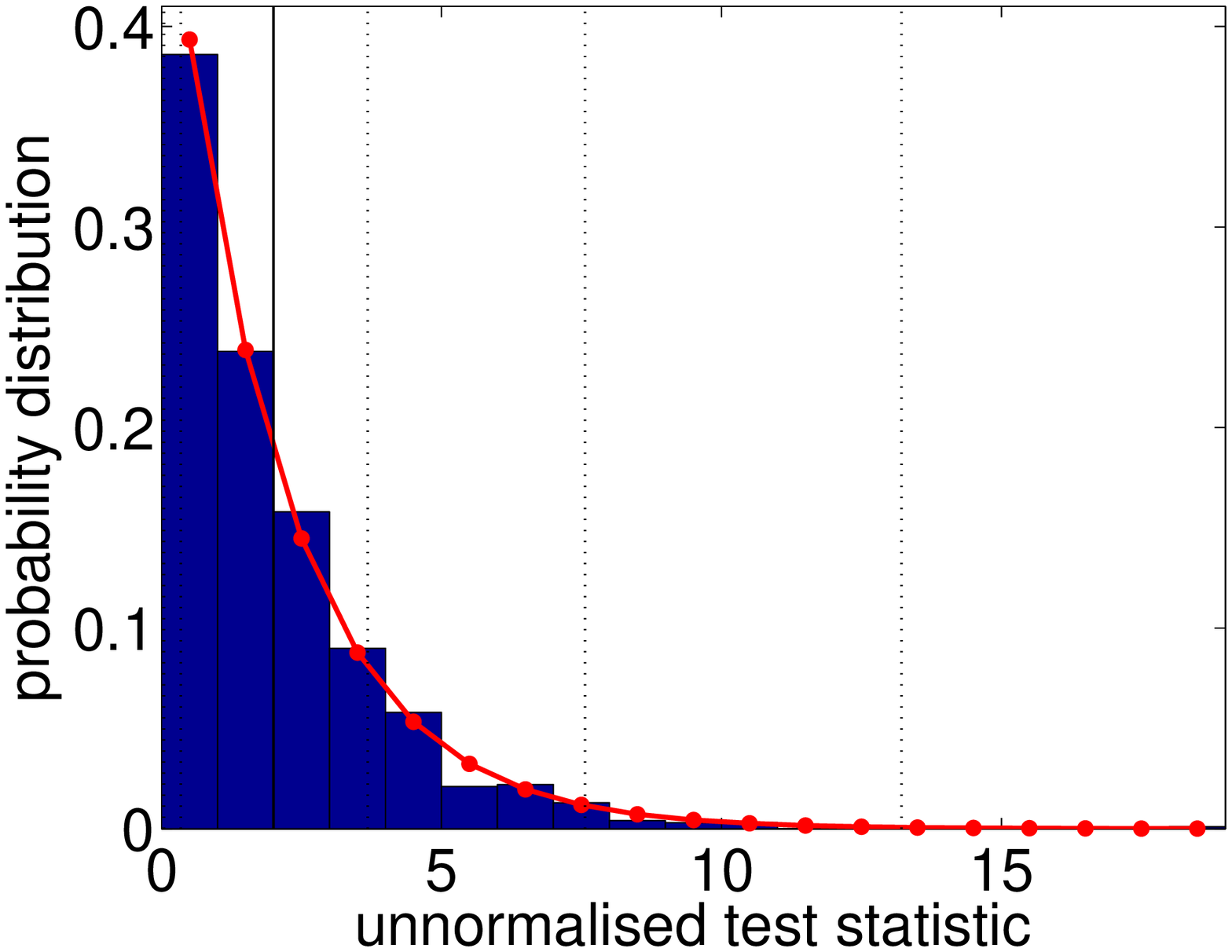} & \hspace{5mm} &
\includegraphics[width=45mm]{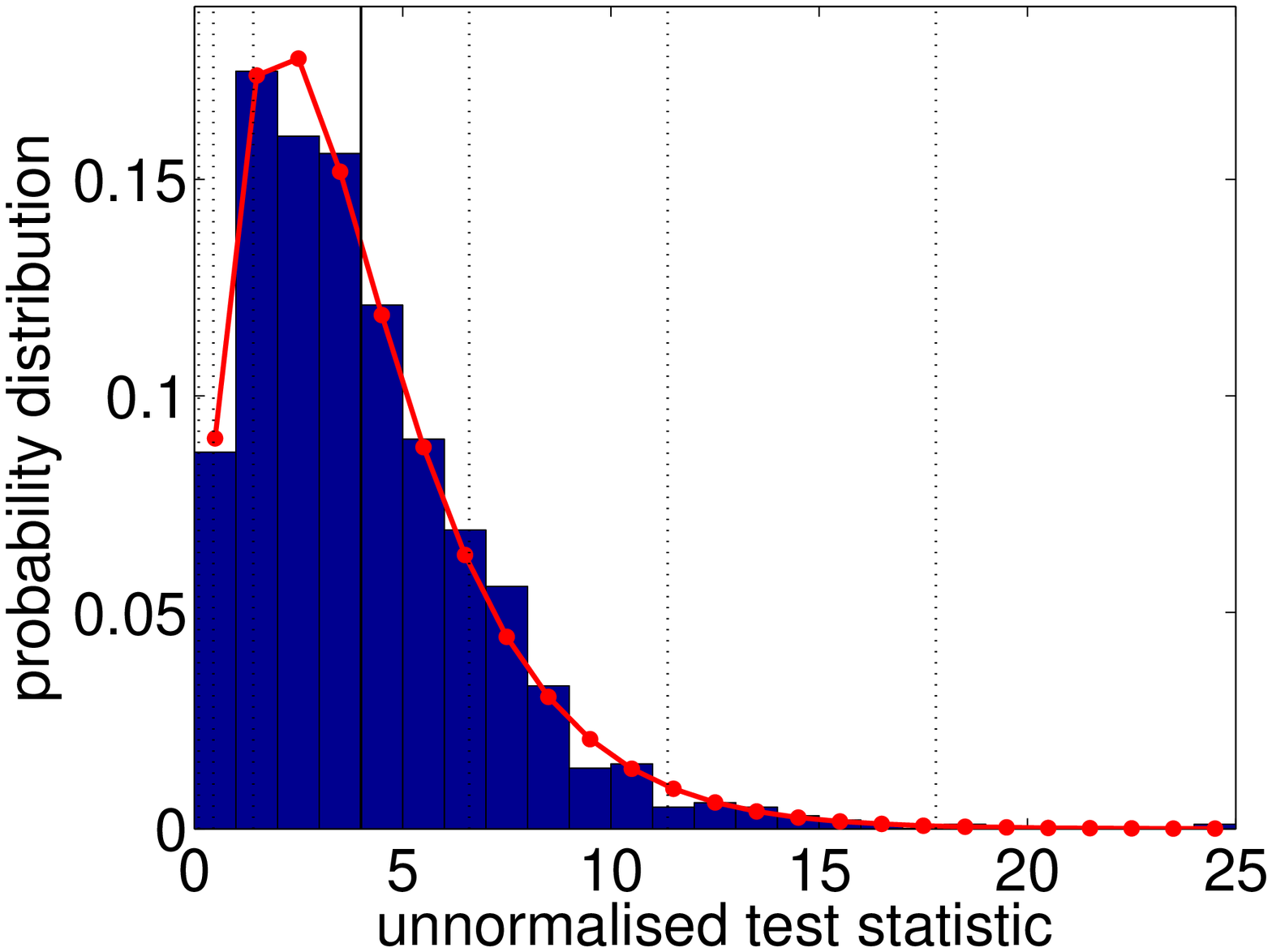} & \hspace{5mm} &
\includegraphics[width=45mm]{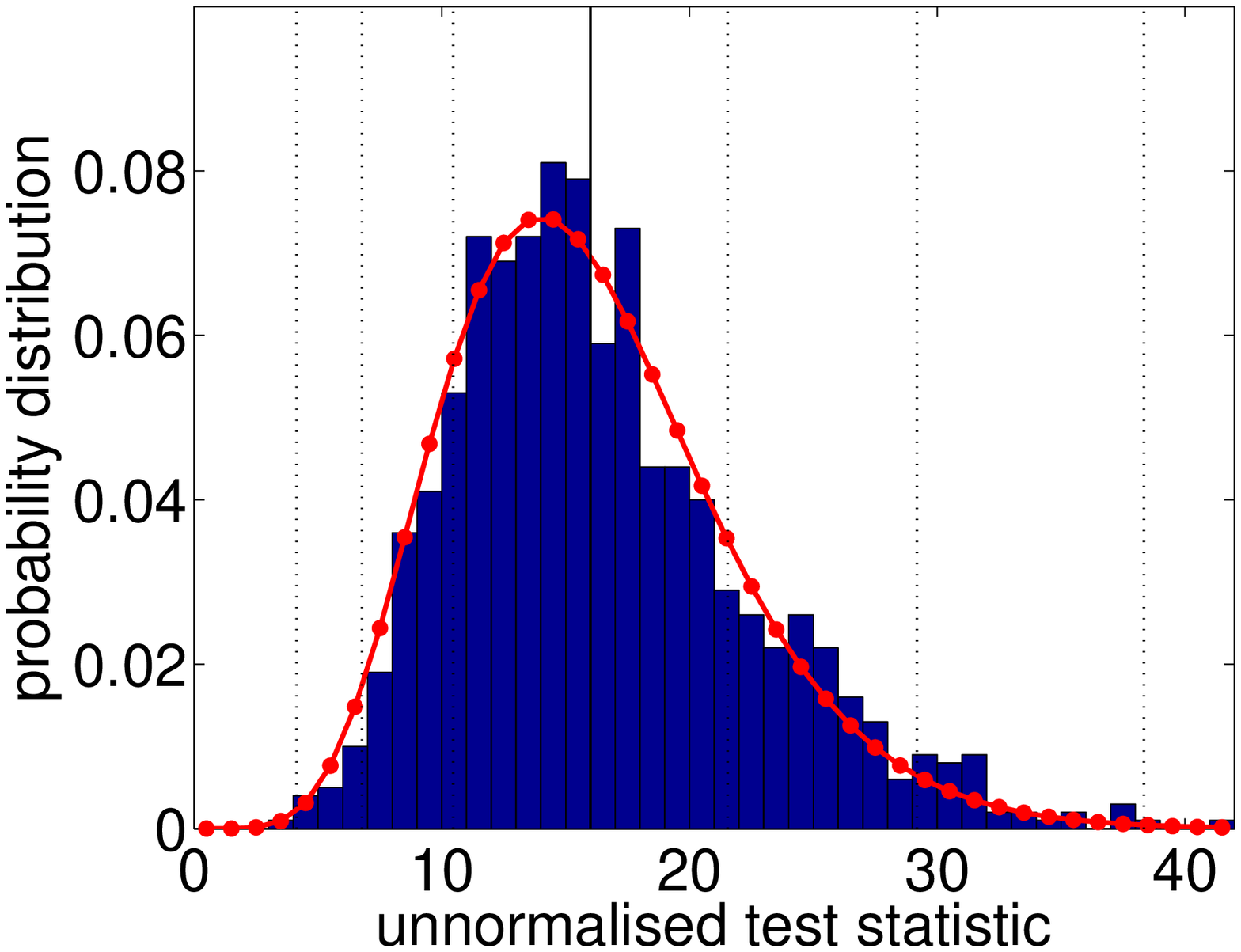} \\
\vspace{+2mm} (a) && (b) && (c) \\
\includegraphics[width=45mm]{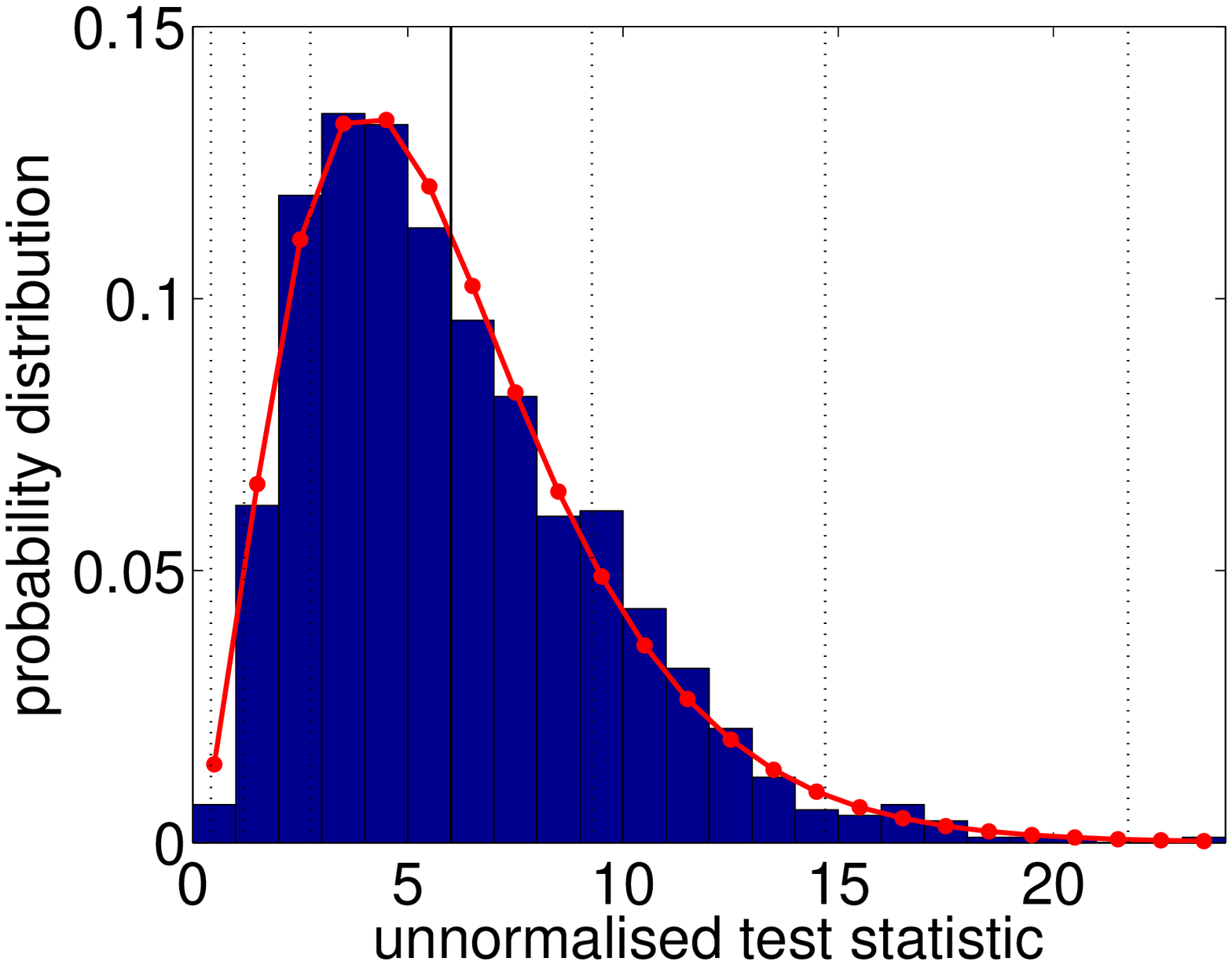} & \hspace{5mm} &
\includegraphics[width=45mm]{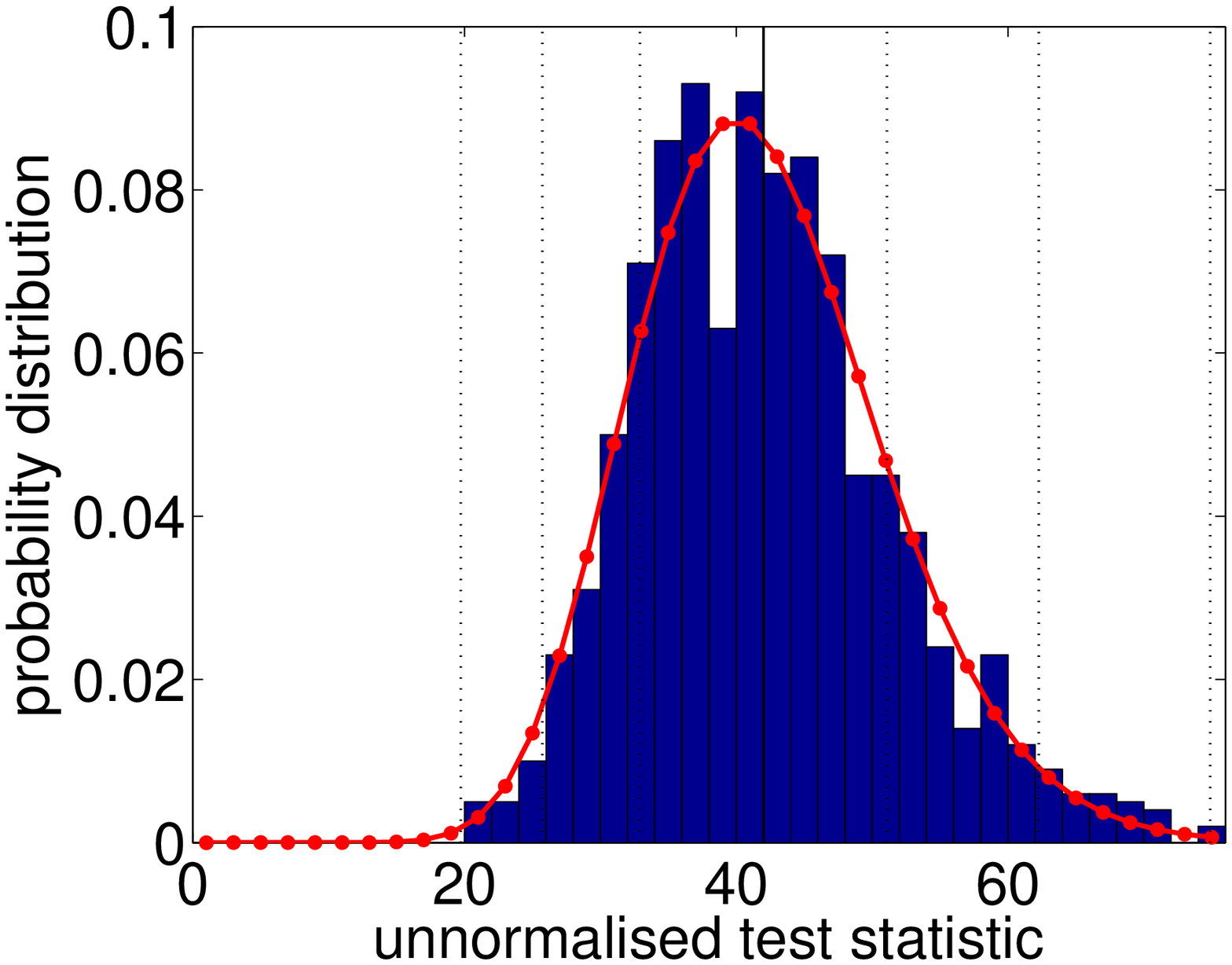} & \hspace{5mm} &
\includegraphics[width=45mm]{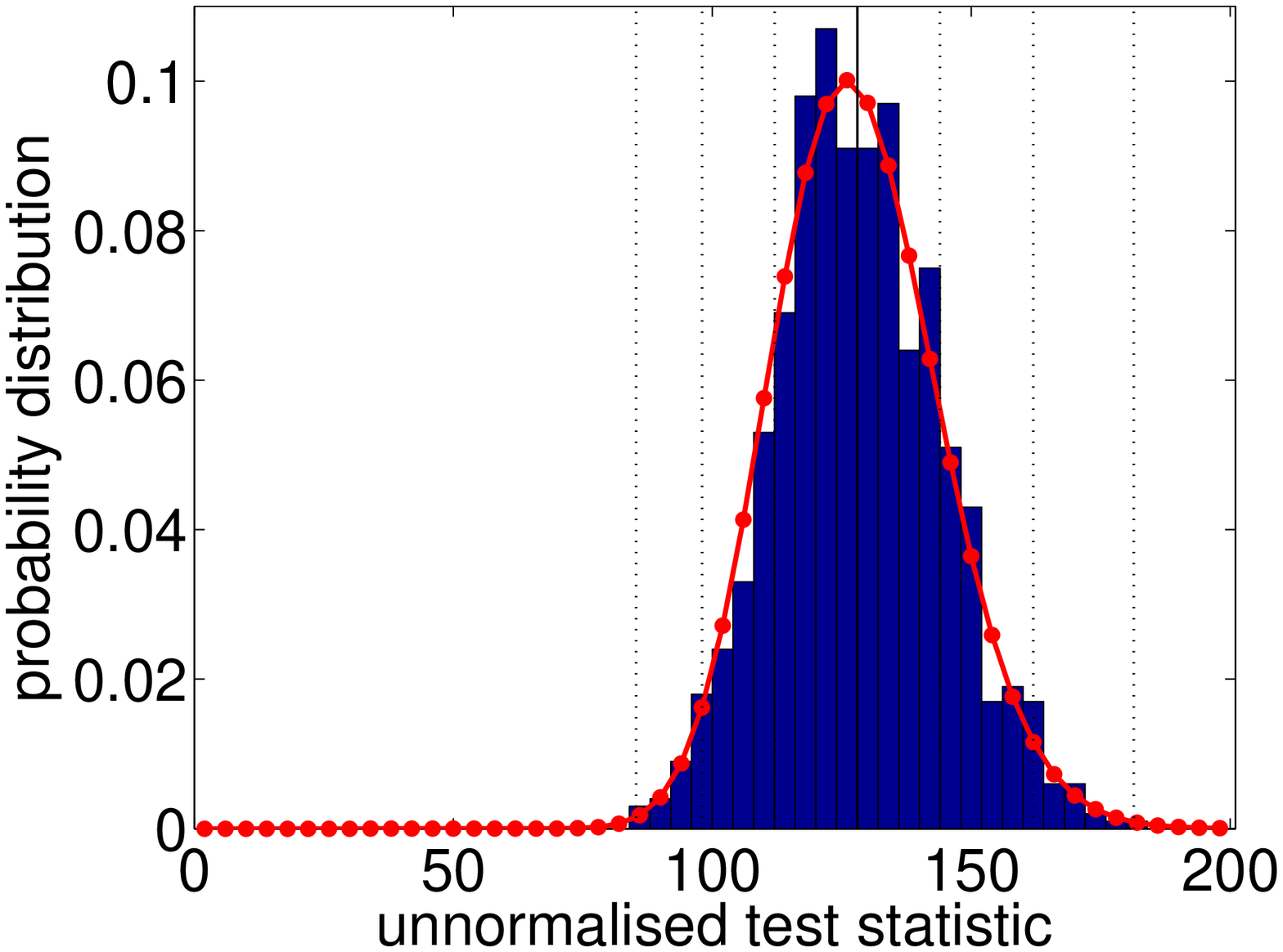} \\
(d) && (e) && (f) \\
\end{tabular}\end{center}
\caption{Comparison of measured distributions of the unnormalised $X^2$ test statistic from numerical tomographies with $\chi^2$ distributions with $M-d^2$ DOFs (red curves).  Simulations were performed using randomly chosen Werner-like mixed states of the form $\rho_t = p\ket{\psi}\bra{\psi} + (1-p) I_d/d$, where $\ket{\psi}$ is an arbitrary pure state randomly chosen from the Haar measure and $p$ was randomly chosen to lie between 1/3 and 2/3.  The example systems shown correspond to: (a-c) single-qubit states with 6, 8 and 20 measurements, respectively; single-qudit states with (d) $d=3$ and $M=15$ and (e) $d=7$ and $M=91$; and (f) two-qubit ($d=4$) states with 12 measurements per qubit ($M=144$).}
\label{fig-penalty-function-dist-mixed}
\end{figure}

\begin{figure}
\begin{center}
\includegraphics[width=90mm]{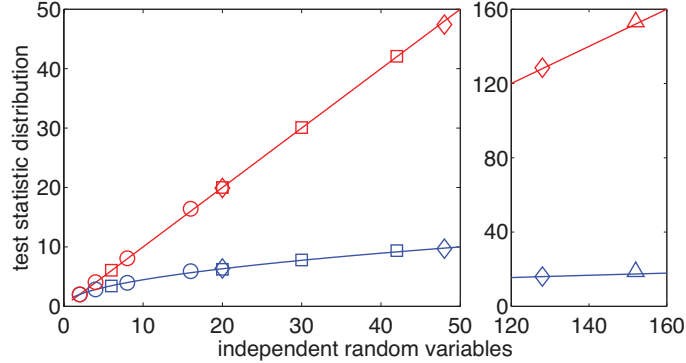}
\end{center}
\caption{Mean (red) and standard deviation (blue) of the measured $X^2$ distribution for numerical tomography simulations performed using randomly chosen mixed states (see also Fig.~\ref{fig-penalty-function-dist-mixed}), plotted as a function of the number of DOFs in the measured data, given an arbitrary full-rank mixed state.  The lines show the expected mean, $M-d^2$, and standard deviation, $\sqrt{2(M-d^2)}$, for the appropriate $\chi^2$ distribution.  The circles ($\circ$) show single-qubit simulations with 6, 8, 12 and 20 measurement settings; the diamonds ($\Diamond$) show two-qubit simulations with 6, 8 and 12 measurements per qubit ($M=36$, $M=64$ and $M=144$); the triangle ($\bigtriangleup$) shows three-qubit simulations with 6 measurements per qubit ($M=216$); and the squares ($\Box$) show single-qudit simulations for $d=3$, 5, 6 and 7, with $M=15$, 45, 66 and 91, respectively.  The error bars are all smaller than the size of the plotted points and the data agrees well with the theoretical prediction within reasonable margins set by those errors.  All of the simulations used 1000 randomly chosen mixed states, except for the three-qubit system, using 200 states.}
\label{fig-penalty-function-trends-mixed}
\end{figure}

To test the usefulness of the reconstruction quality defined above, it is first necessary to check whether $X^2$ does conform to a $\chi^2$ distribution in typical experimental scenarios.  Figure~\ref{fig-penalty-function-dist-mixed} shows examples of a range of numerical tomography simulations performed with different system sizes and measurement sets.  Each case considers 1000 random Werner-like mixed states of the form $\rho_t = p\ket{\psi}\bra{\psi} + (1-p) I_d/d$, where $\ket{\psi}$ is a arbitrary pure state randomly chosen from the Haar measure and $p$ was randomly chosen to lie between 1/3 and 2/3.  The six plots show histograms of the measured $X^2$ test statistic distributions for typical examples, exhibiting good agreement with the probability density function for a $\chi^2$ distribution with the $M-d^2$ DOFs expected given $d^2$ fitting parameters for an arbitrary density matrix (also plotted).  The means and standard deviations of the $X^2$ distributions for all systems studied ($d=2$ to $d=8$ and $M=6$ to $M=216$) are plotted in Fig.~\ref{fig-penalty-function-trends-mixed} against the number of independent DOFs, $M-d^2$.  The lines show the expected mean and standard deviation of $M-d^2$ and $\sqrt{2(M-d^2)}$, respectively.  Throughout this paper, I will consider example measurement sets based on the platonic solids for qubit-based systems~\cite{deBurghMD2008}, given by: a tetrahedron (4 measurements, minimally complete), a cube (6 measurements, over-complete), an octahedron (8 measurements), a dodecahedron (12 measurements), and an icosahedron (20 measurements).  For qudit-based systems, I will use an overcomplete set which is a generalised form of the cube set for qubits, and which includes one measurement for each of the computational basis states and four measurements to probe each of the density matrix coherences (two real and two imaginary two-state superpositions with opposite signs)~\cite{LangfordNK2004}.

\begin{figure}
\begin{center}\begin{tabular}{ccc}
(a) & (b) & (c) \\
\includegraphics[width=45mm]{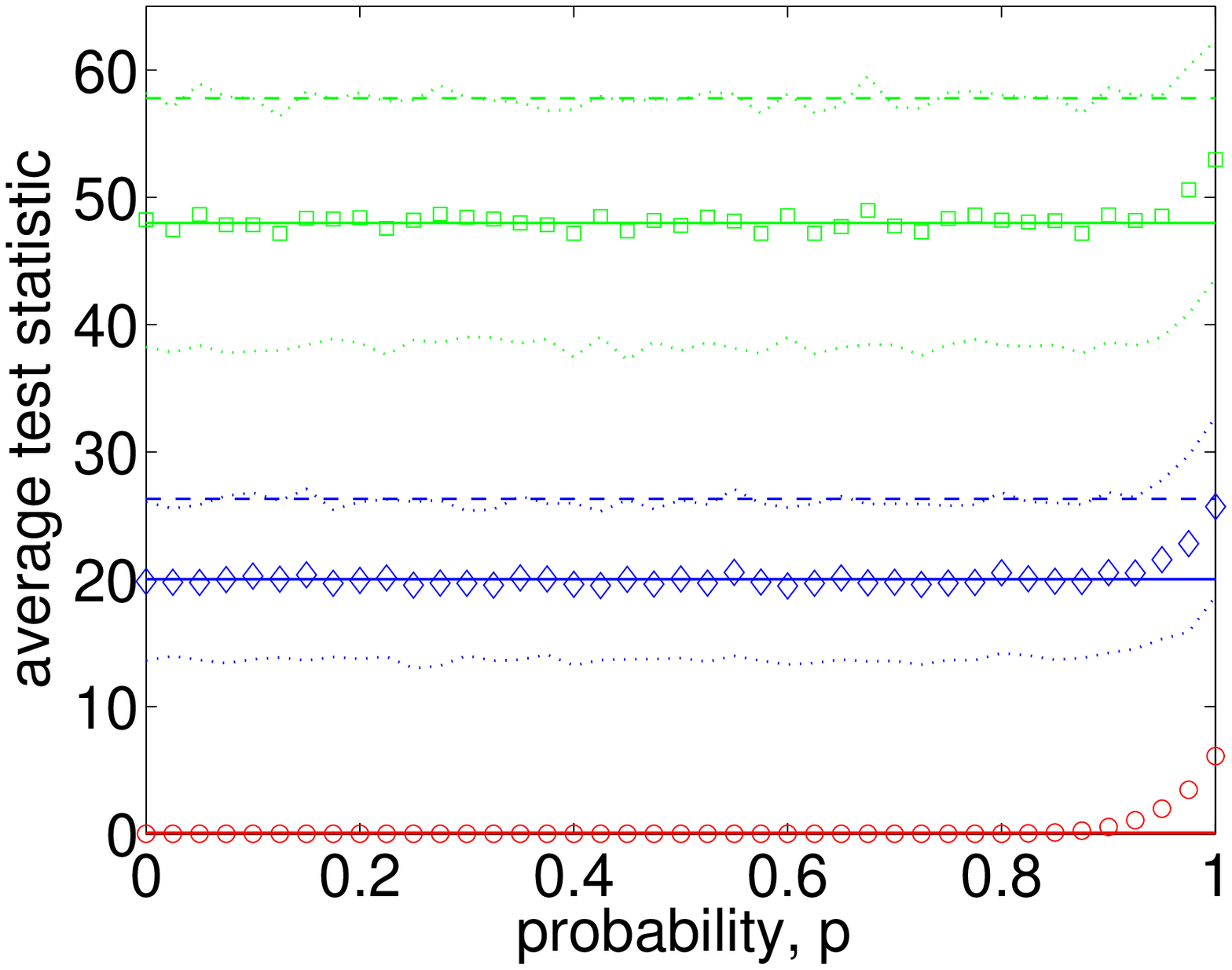} &
\includegraphics[width=45mm]{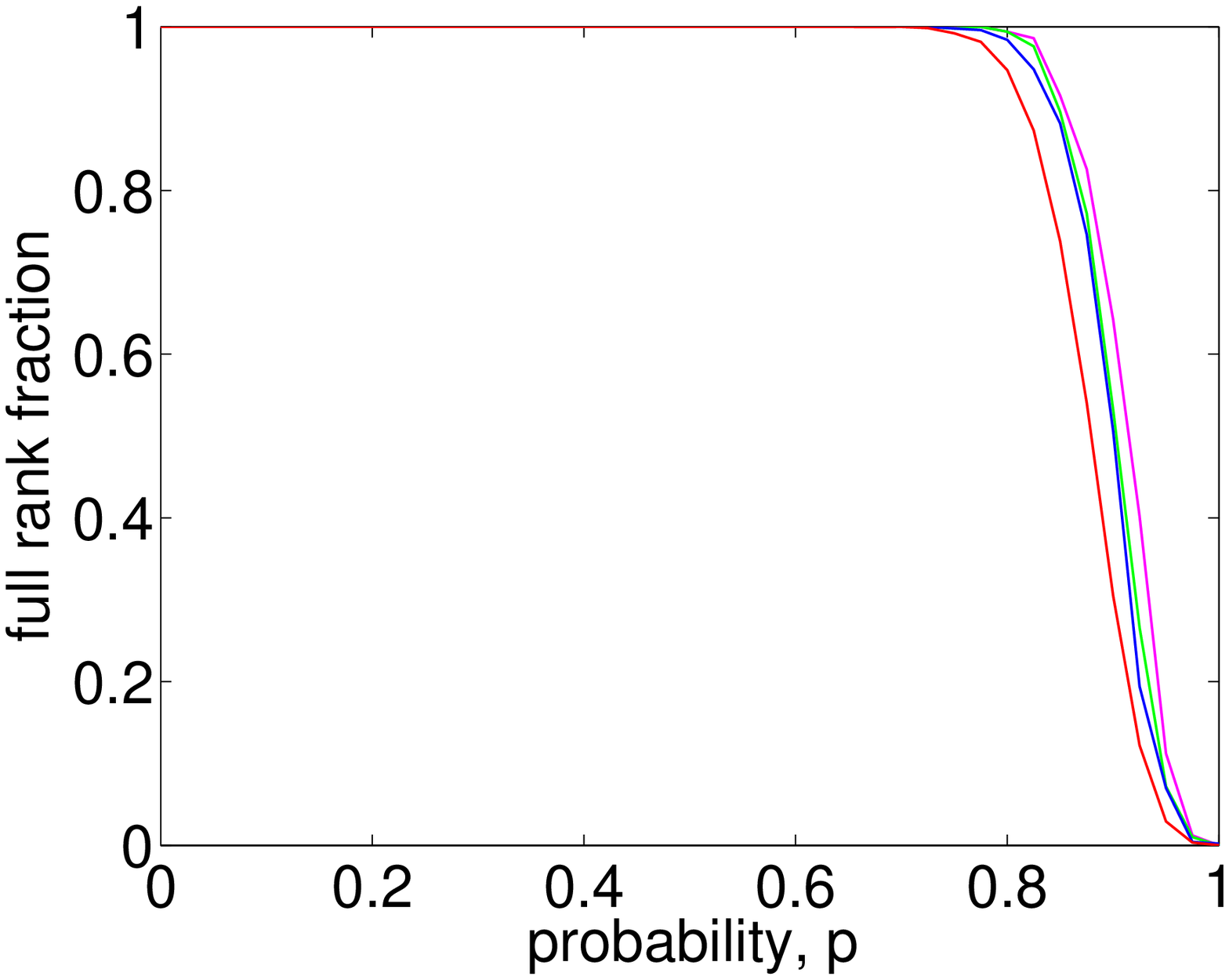} &
\includegraphics[width=45mm]{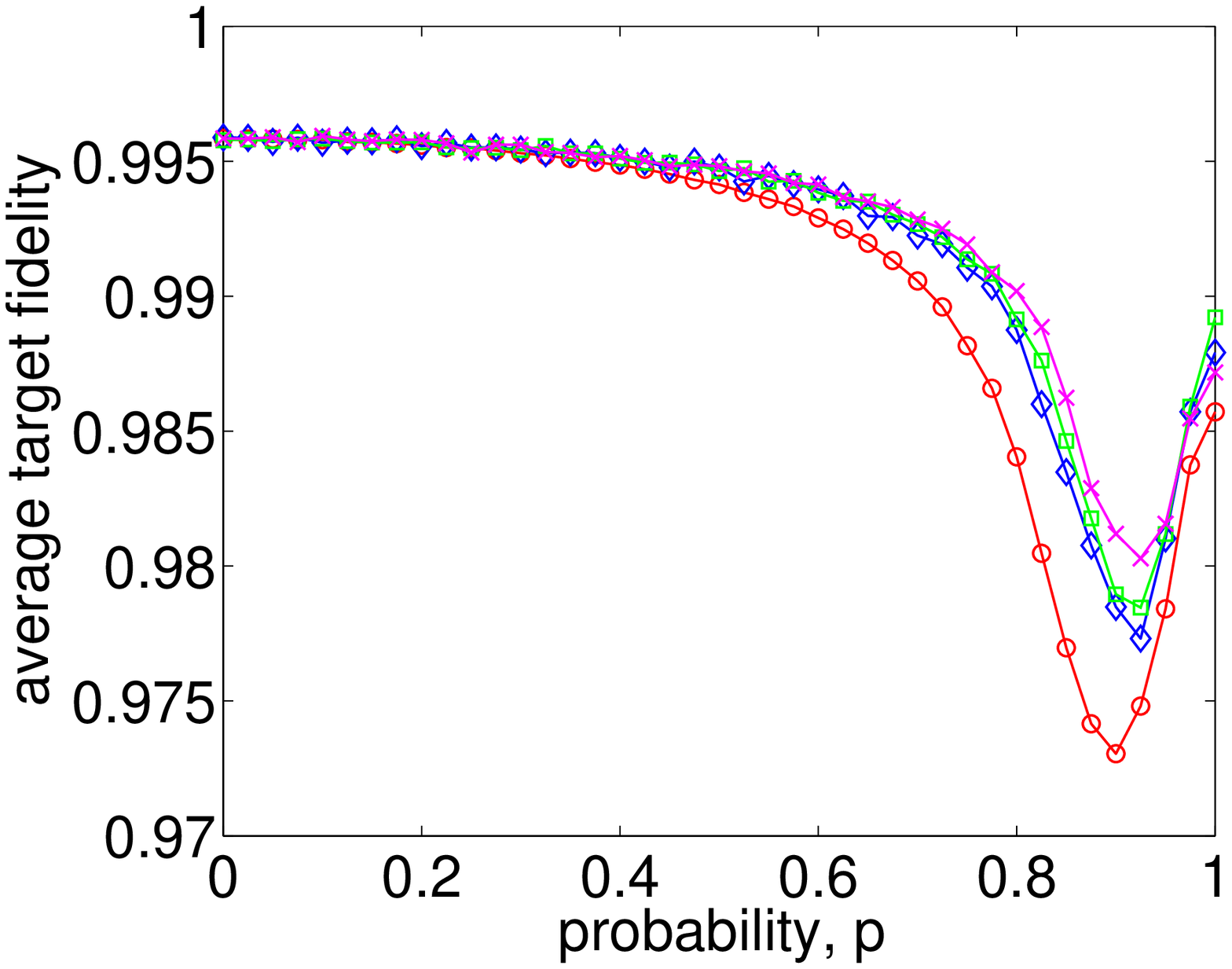}
\end{tabular} \\
\begin{tabular}{cc}
(d) & (e) \\
\includegraphics[width=45mm]{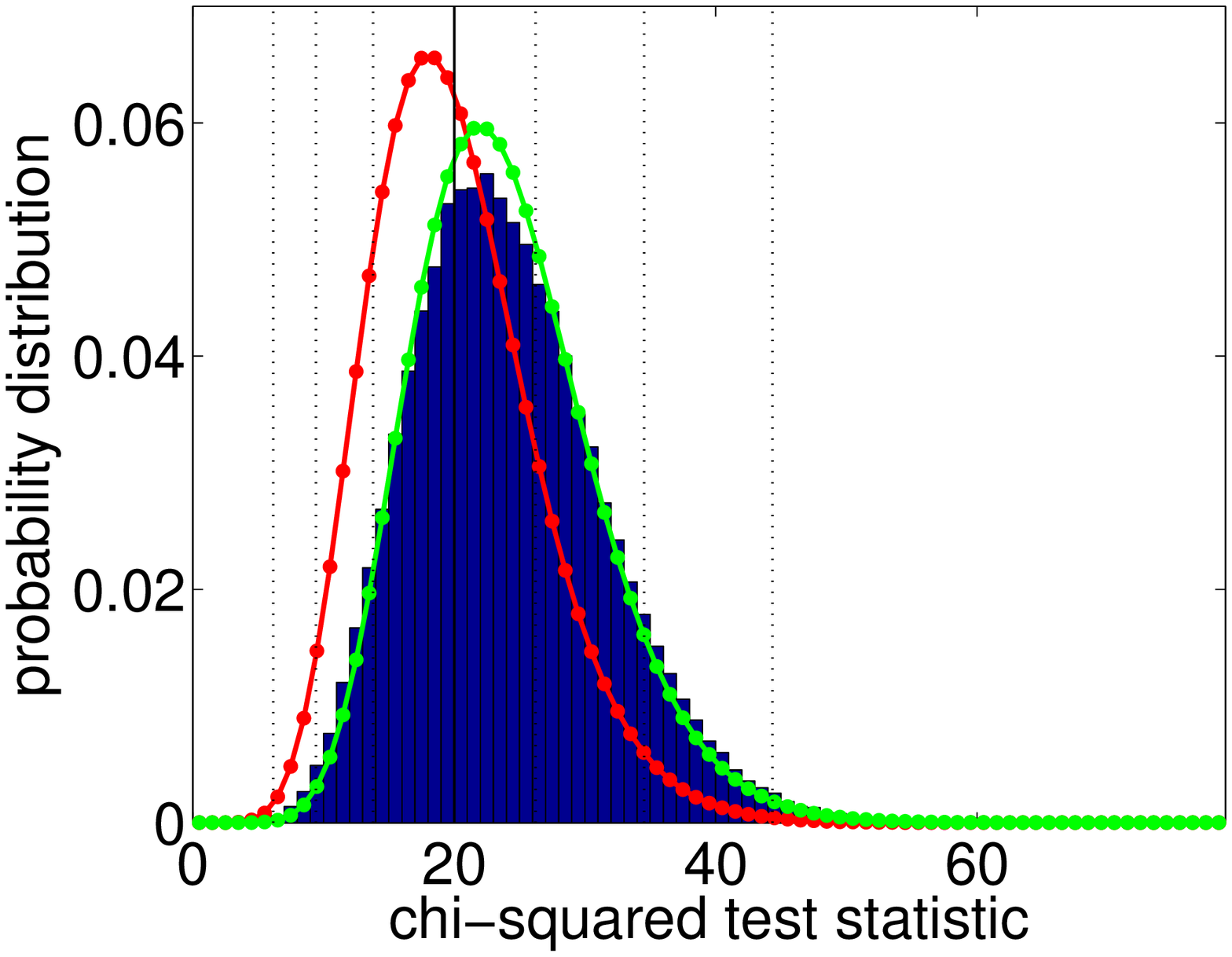} &
\includegraphics[width=45mm]{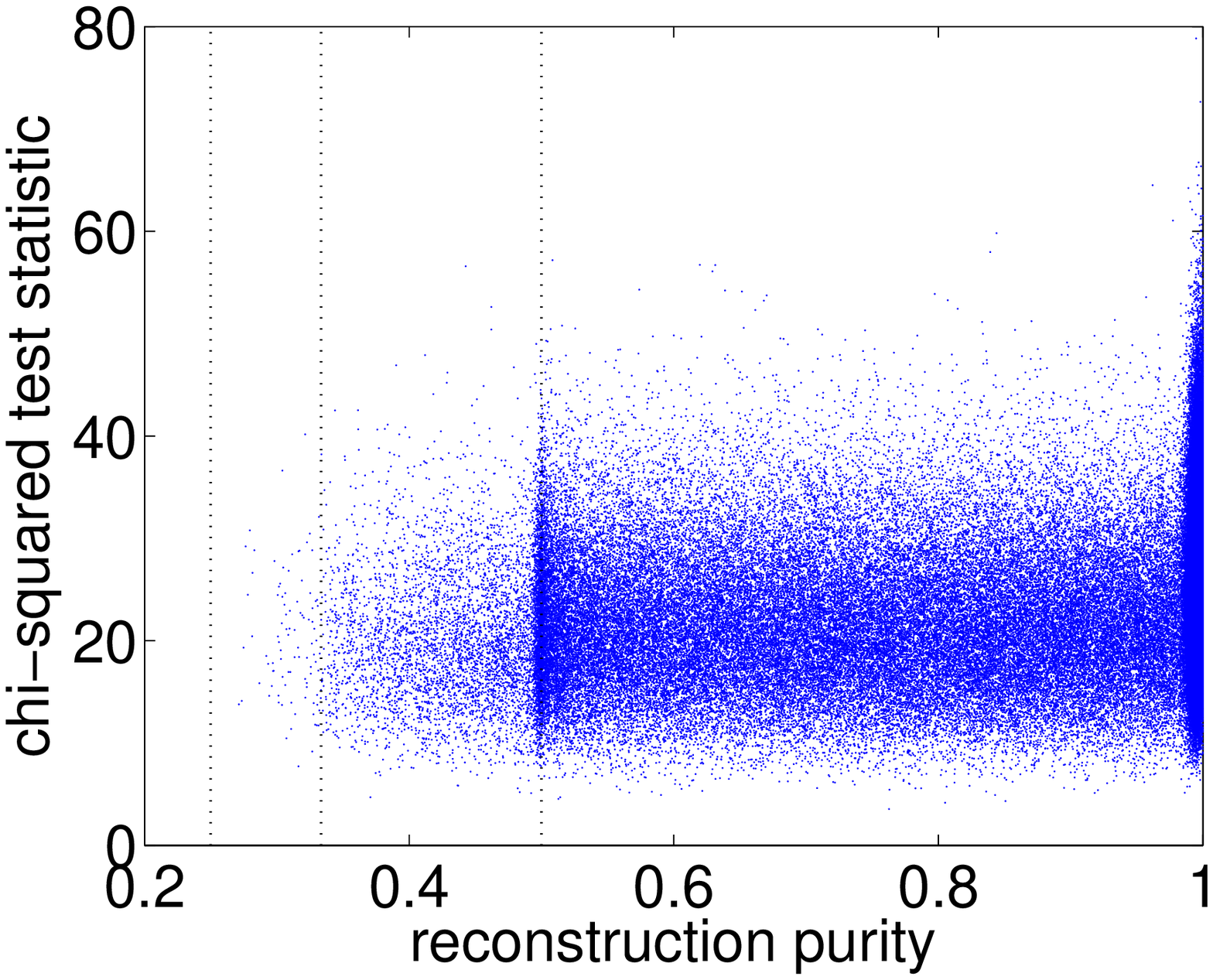}
\end{tabular}\end{center}
\caption{Simulations illustrating the effect of mixture on $X^2$ in two-qubit tomography. (a--c) Werner states~\cite{WernerRF1989} of the form $\rho_t = p\ket{\psi}\bra{\psi} + (1-p) I_4/4$ ($I_4$ is the identity operator in four dimensions and $\ket{\psi}$ is a maximally entangled state) studied as a function of $p$.  For each value of $p$, randomly chosen states are used to generate Poissonian-distributed noisy data sets for different platonic solid measurement sets~\cite{deBurghMD2008}: tetrahedron (minimally complete, red: {\color[rgb]{1,0,0}---$\circ$---\color[rgb]{1,0,0}}), cube (over-complete, blue: {\color[rgb]{0,0,1}---$\Diamond$---\color[rgb]{0,0,1}}), octahedron (green: {\color[rgb]{0,1,0}---$\Box$---\color[rgb]{0,1,0}}), and dodecahedron (magenta: {\color[rgb]{1,0,1}---$\times$---\color[rgb]{1,0,1}}); which are then reconstructed using maximum likelihood estimation.  Each point represents an average over 500 random states, except in the four-measurement set data, which averages over 7500 states per point.  The graphs show (a) the average chi-squared test statistic for the noisy reconstructed states (the much-larger-valued twelve-measurement set exhibits the same trend, but is omitted in order to show better detail with the other sets); and (b) the fraction of reconstructed states which are full rank; and (c) the average fidelity of the noisy reconstructed states with the known target state.  In (a), the solid lines show the expected value of the test statistic, assuming $M-d^2$ DOFs and the dashed lines show the expected position of the first standard deviation higher.  The dotted lines show the width of the observed distributions (because they are skewed, the lines show the $\chi^2$-function positions which correspond to the integrated probabilities for the $\pm \sigma$ positions of a symmetric Gaussian distribution).  (d--e) 150,000 randomly chosen, mixed two-qubit state reconstructions (6 measurements per qubit) with different numbers of nonzero eigenvalues. (d) Comparison of the observed distribution of $X^2$ with the theoretical $\chi^2$ probability distribution for $M-d^2=20$ DOFs (red curve) and an optimal $\chi^2$ distributions with a fit to the number of DOFs over arbitrary real numbers (green curve:  $M{-}c \sim 23.97$).  Vertical lines show the expected means and standard deviation positions.  (e) $X^2$ vs purity ($\Tr{\varrho^2}$) of the reconstructed states.  Dotted lines at $\Tr{\varrho^2}=1/4$, 1/3 and 1/2 show the minimal purities for states with 4, 3 and 2 nonzero eigenvalues, respectively.}
\label{fig-penalty-function-mixture-trends}
\end{figure}

The above simulations were all carried out using randomly chosen mixed states which always resulted in reconstructed density matrices with full rank ($d$ eigenvalues), where the number of DOFs can be expected to match the value predicted using the full $d^2$ free parameters of an arbitrary mixed state as constraints.  Figure~\ref{fig-penalty-function-mixture-trends}, on the other hand, shows the results of numerically simulated tomographies of randomly chosen Werner states (mixtures of maximally entangled states with the maximally mixed state) as a function of the population of the pure, maximally entangled component of the state.  These random states were used to generate noisy data for different measurement sets based on tensor products of single-qubit, platonic-solid measurements, but with the same total tomography time for each (i.e., correspondingly less time per setting for the larger sets).  Figure~\ref{fig-penalty-function-mixture-trends}(a) demonstrates again that, for higher levels of mixture, the observed distribution of $X^2$ agrees well with the corresponding $\chi^2$ distribution with $M-d^2$ DOFs.  For higher purities, however, the means of the observed distributions begin to deviate substantially above the expected values, suggestive of a decrease in the number of constraints imposed by the fit of free parameters in the density matrix.  This agrees at least qualitatively with what would be expected for higher purities, given that a pure state contains substantially fewer free parameters than a full-rank mixed state.  It is further supported by Fig.~\ref{fig-penalty-function-mixture-trends}(b), which shows that the increase in $X^2$ follows the decrease in fraction of reconstructed states of full rank (containing no nonzero eigenvalues).

But how significant is this discrepancy?  Figure~\ref{fig-penalty-function-mixture-trends}(a) shows that, for the cube setting (six measurements per qubit), the observed mean moves almost one standard deviation higher than the expected value based on $d^2$ constraints.  The significance of such a shift can be seen by considering how the chi-squared test statistic is designed to be used.  Ultimately, we wish to be able to detect the presence of unexpected noise in the raw tomography data, to determine whether the reconstruction is in fact reliable.  Assuming that the increase in mean results from a straightforward increase in DOFs, a na\"ive calculation shows that a shift of one standard deviation would result in around $23\%$ of excess-noise-free data sets being diagnosed as excess-noise-affected at the 95\% confidence level, instead of the 5\% expected from the confidence definition.  In reality, this assumption is too simplistic, as Figs~\ref{fig-penalty-function-mixture-trends}(d) and~(e) explore in more detail.

Figure~\ref{fig-penalty-function-mixture-trends}(d) shows the observed distribution of the chi-squared test statistic, $X^2$, for 150,000 randomly chosen, mixed target two-qubit states.  The target mixed states were chosen by first generating a random number of random eigenvalues in a way which biased the states towards the boundaries of the physical state space, to ensure that there were reconstructed states of every rank, and then assigning to each nonzero eigenvalue a pure eigenvector uniformly distributed according to the Haar measure (made orthogonal via Gram-Schmidt reduction).  Figure~\ref{fig-penalty-function-mixture-trends}(e) provides a qualitative verification that this method does produce mixed states throughout the state space.  These states were then used to generate noisy measurement sets (with Poissonian noise) with 6 measurements per qubit ($M=36$).  The red curve shows the $\chi^2$ distribution for $M{-}d^2 = 20$ DOFs and the green curve shows the optimal $\chi^2$ distribution found by fitting the number of independent parameters ($M{-}c \sim 23.97$).  Using, somewhat self-referentially, a standard chi-squared test to determine whether the observed histogram could be plausibly obtained from either of the theoretical $\chi^2$ distributions, the calculated $p$-value was 0 in both cases (to within machine precision).  Thus, unlike the earlier example of full-rank mixed states illustrated in Figs~\ref{fig-penalty-function-dist-mixed} and~\ref{fig-penalty-function-trends-mixed}, the test statistic for an arbitrary matrix cannot simply be characterised in terms of a $\chi^2$ distribution with a single well-defined number of free parameters.  In fact, it is known that hypothesis testing statistics do not generally, in the presence of inequality constraints, asymptote towards a simple $\chi^2$ distribution, but instead often towards a mixture of $\chi^2$ distributions with different DOFs, known as a chi-bar-squared ($\bar{\chi}^2$) distribution~\cite{ShapiroA1988, ElBarmiH1999, SilvapulleMJ1996}.  This can be seen qualitatively from Fig.~\ref{fig-penalty-function-mixture-trends}(e), where several distributions in $X^2$ with different means and widths are clearly visible at different purities.

Applying the same reasoning as above, if these results were analysed using the assumption of $d^2$ free parameters in the density matrix, around 16\% (4.8\%) of the reconstructed states would be diagnosed as affected by excess noise at the 95\% (99\%) confidence level, although the raw counts do only contain known statistical noise.  Using this approach would therefore substantially overestimate the presence of noise when trying to diagnose systematic errors, severely undermining the validity of the chi-squared test statistic as a quantitative diagnostic in quantum tomography.  Similar results would also be observed when using alternative statistical tests, such as the likelihood-ratio tests proposed in Ref.~\cite{MoroderT2012}, which are known to result in $\bar{\chi}^2$ limiting distributions~\cite{ElBarmiH1995,ElBarmiH1999}.

Returning briefly to Fig.~\ref{fig-penalty-function-mixture-trends}(a), note that the octahedron setting (eight measurements per qubit) exhibits a smaller discrepancy by comparison with the standard deviation of the expected $\chi^2$ distribution ($2\kappa$).  This results from the fact that the maximum range of DOFs is determined only by the range of constraints ($2d-1$ to $d^2$), producing a smaller relative discrepancy for larger measurement sets.  Unfortunately, it still scales badly as a function of system size for multiqudit systems.  A rough calculation for larger numbers of qudits shows that this effect becomes exponentially more pronounced with increasing system size unless the number of measurements per qudit is greater than $d^4$, where in this case $d$ is the single-qudit dimensionality (i.e. $>16$ for qubits).  Note that this holds for both photonic systems as well as atomic-like systems with near-deterministic measurements.

It is therefore critical to develop a method for accurately dealing with the reduced number of independent DOFs in tomographic reconstructions near the physical state boundary if it is going to be possible to diagnose systematic noise in a rigorous quantitative way.

\subsubsection{Application example: a note on measurement sets and completeness}

Despite the quantitative discrepancy in the observed value of the chi-squared test statistic at high purities, it can already provide useful qualitative information.

Figure~\ref{fig-penalty-function-mixture-trends}(c) shows the average target-state fidelity, demonstrating that the over-complete sets perform somewhat better than the minimally complete, four-measurement set at larger purities.  This supports the conclusions of Ref.~\cite{deBurghMD2008} that, despite a larger number of measurements and concomitantly shorter integration times per measurement, over-complete measurement sets offer advantages over even optimal, minimal measurement sets~\cite{LingA2006}.  Indeed, the minimally complete measurement set only seems to perform comparably well to the over-complete sets at higher levels of mixture where, particularly for these Werner states which smoothly increase in symmetry as the mixture increases, the measured data contains almost no information, in the sense that all counts should be approximately equal, irrespective of what measurement is being made.  By contrast, there are substantial improvements for over-complete measurements for states near the boundary of physical states, which is generally the region of most interest for QIP contexts.  This can be understood fairly easily by looking at the $X^2$ test statistic.

Figure~\ref{fig-penalty-function-mixture-trends}(a) shows that, provided the state is sufficiently far from the state-space boundary relative to the measurement noise, the minimally complete tomography produces a \emph{perfect} fit to the data every time, even when the data is known to contain noise.  Except near the pure-state boundary where the physicality constraints can play a role, the minimally complete measurement set is completely unable to distinguish between quantum state and noise, and the tomography faithfully reconstructs the noise instead of the underlying state.  The over-complete measurement set, however, contains redundant information which allows the tomographic optimisation to diagnose and reject some of the noise present in the data.  While inevitably resulting in some mismatch between reconstruction and data ($X^2 {\neq} 0$), this pulls the reconstruction towards the underlying state, resulting in a reconstruction which is at least as accurate as the minimal tomography and generally more so---even though the integration time at each setting is substantially less (less than half in this example).  Furthermore, it is only by taking over-complete measurements that it is even conceivable to use the data to check the validity of the underlying model.  But this should not really be surprising.  When asking students in an undergraduate laboratory to measure a quantity related to the gradient of a function, we never ask them to take two data points and draw a line between them.  We expect them to measure a series of data points and use a numerical fit to all the data and our expectations should probably be no different for quantum tomography.

Essentially, there are two guiding principles for choosing tomographic measurement sets: 1) using over-complete measurement sets with redundancy of information will generally perform better than minimally complete sets, provided they are equivalently symmetric; and 2) the accuracy of tomography will increase as coverage of the Hilbert space increases, but for multiple qubits, this would generally require the use of entangled measurements which may therefore introduce other challenges that could counteract potential improvements.

It is worth noting that determining exactly what constitutes a minimally complete set of measurements can be nontrivial, requiring a detailed understanding of the experimental specifics of the measurement apparatus.  For example, for the photonic systems described above, where each measurement setting probes a single projector, four settings per qubit constitutes a minimal set (the final parameter is required for normalisation purposes, so the minimal set has $d^q$ elements, where $q$ is the number of qubits, not the usual $d^{q}{-}1$).  In ion-trapping tomography, however, where the measurements are essentially deterministic and the number of system copies used is therefore fixed and well-defined, it is typical to use three measurement settings per qubit.  For a single qubit, this gives six measured probabilities, but still only three independent parameters.  For two qubits, however, the four probabilities for each setting include three independent parameters, corresponding to 27 independent parameters for all 9 settings, and this already provides a substantial amount of measurement redundancy~\cite{MoroderT2012}.  So although the standard three settings provide a minimally complete set for one qubit, they produce over-complete sets for two or more qubits.

Interestingly, this observation explains trends illustrated in Ref.~\cite{deBurghMD2008} as well as Fig.~\ref{fig-penalty-function-mixture-trends}, which justify the guiding principles described above.  One-qubit simulations show that using more measurements generally improves tomographic accuracy, resulting in some qualitative sense from greater coverage of the underlying Hilbert space.  By contrast, when using separable measurement sets constructed from tensor products of single-qubit measurements, simulations with two or more qubits generally show some improvement between a minimally complete set and an over-complete set, but that little further advantage is gained by increasing the number of single-qubit measurements beyond what gives rise to an over-complete set.  For example, unlike the one-qubit case, in the atomic-like systems, the standard cube-based measurement is already near-optimal for two or more qubits~\cite{deBurghMD2008}.  This saturation of further improvement can be understood by noting that separable tensor-product projectors are a measure-zero subset of the space of all pure measurements~\cite{BengtssonZyczkowski}.  That is, while it is possible for a set of separable measurements to span a multiqubit Hilbert space, provided the measurement set is over-complete, adding more separable measurements is unlikely to substantively increase the coverage of the encompassing space, which almost entirely consists of entangled states.  Thus, for multiple qubits, further significant improvements beyond the cube-based tomographic sets are only likely to be achieved using at least some entangled states (see, e.g., Ref.~\cite{AdamsonRBA2010}).

\subsection{Calculating degrees of freedom and the reduced reconstruction quality}

Figure~\ref{fig-penalty-function-mixture-trends} provides an indication of the complexities that arise when trying to understand the effects which inequality physicality constraints impose on the statistics of goodness of fit in tomographic reconstructions, especially as a function of important experimental variables such as state purity.  As illustrated in Fig.~\ref{fig-penalty-function-mixture-trends}(e), such inequality constraints can produce complex distributions incorporating several, distinctly different underlying distributions.  Determining the precise form of this expected distribution in advance is known to be a complex or even intractable problem~\cite{DykstraR1991,StoelRD2006} and, at this stage, more investigation is still required to find the ideal solution for the context of tomography.  In this section, however, I suggest a simple, heuristic way to try and compensate for the different conditions imposed by boundary conditions on different tomographic reconstructions.  The intuition motivating this approach builds on the fact that when the physicality constraints come into play, they tend to result in reconstructions with some zero eigenvalues, thus reducing the number of free parameters in the reconstructed density matrix [see, e.g., Fig.~\ref{fig-penalty-function-mixture-trends}(a)] and in turn also the number of constraints imposed by optimisation.  One might therefore expect that tomographic reconstructions which are more strongly affected by the physicality conditions would tend to arise from distributions characterised by more DOFs.

The most obvious way to work out the number of constraints $c$ introduced by tomographic optimisation is to count the number of independent parameters in an arbitrary density matrix with $l$ nonzero eigenvalues ($1 \le l \le d$).  We will assume that the matrix is unnormalised to allow the calculation of an unknown normalisation parameter during the tomographic reconstruction, but this also makes the calculation easier.  The first eigenvector is an unknown, arbitrary $d$-dimensional pure state, so it is described by $2d-1$ independent parameters.  The next eigenvector, however, must be orthogonal to the first and therefore lies in the $(d-1)$-dimensional subspace which remains when the first eigenvector is removed from the overall state space.  It therefore adds a further $2(d-1)-1$ independent parameters to the density matrix.  Continuing in this manner until the right number of eigenvectors have been added, it is straightforward to calculate the total number of independent parameters, and therefore also the number of constraints to be:
\begin{align}
\label{eq-mixed-state-constraints}
c = \sum_{j=0}^{l-1} \sqbr{2(d-l)-1} = l(2d-l),
\end{align}
which gives $c=2d-1$ for pure states and $c=d^2$ for full-rank mixed states, as required.

\begin{figure}
\begin{center}\begin{tabular}{ccc}
(a) & \hspace{5mm} & (b) \\
\includegraphics[width=45mm]{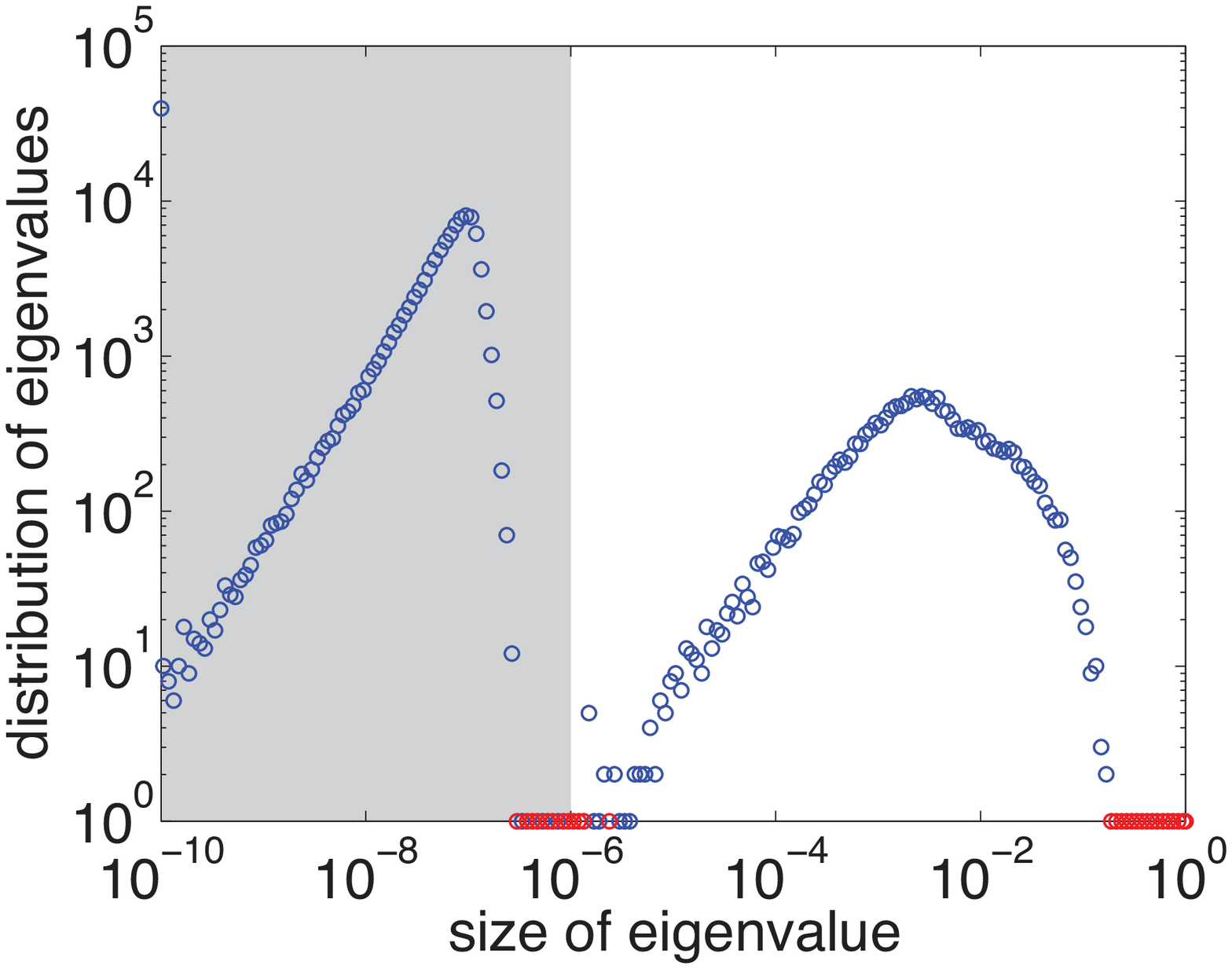} &\hspace{30mm}&
\includegraphics[width=45mm]{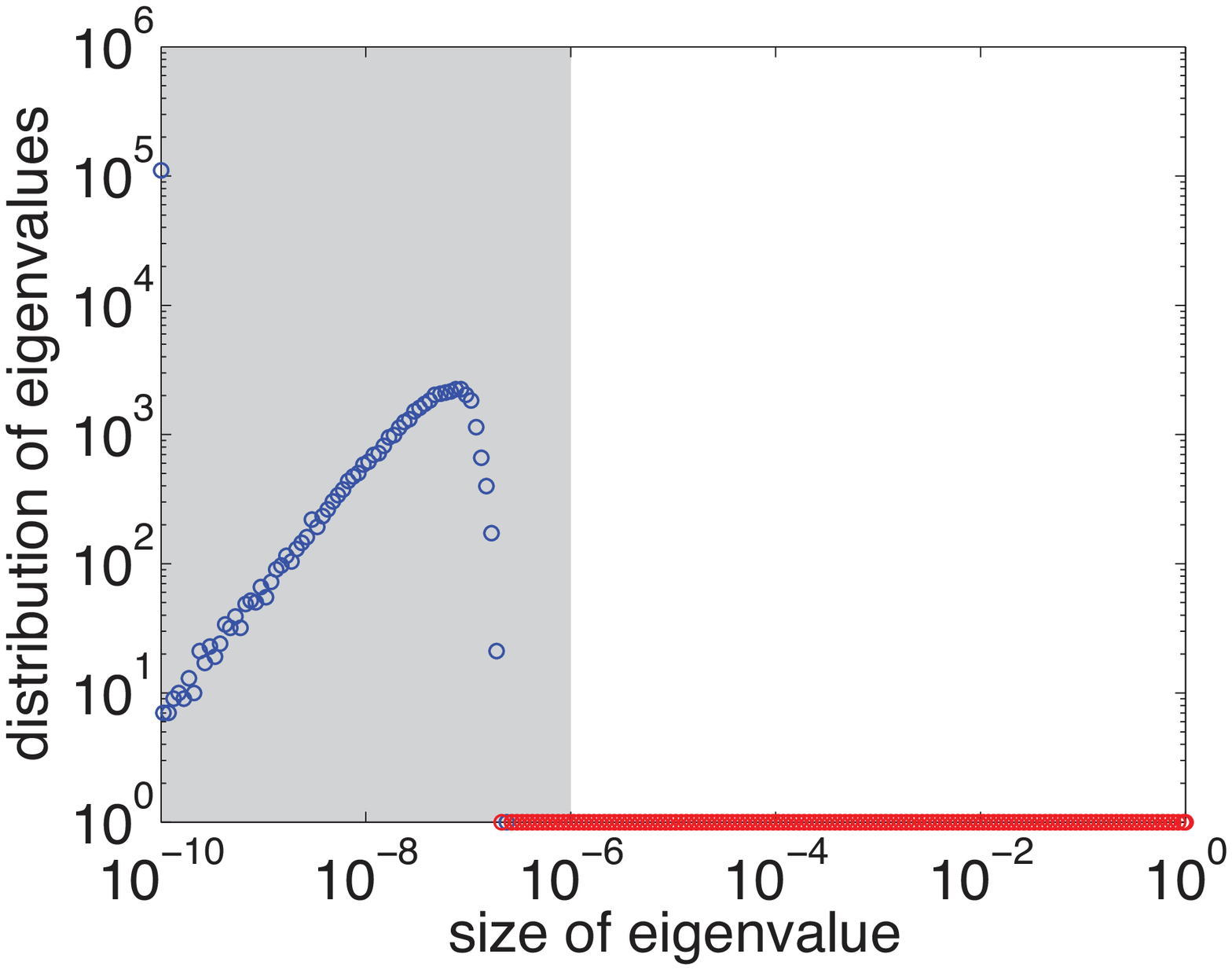}
\end{tabular}\end{center}
\caption{Counting the number of nonzero eigenvalues.  Histograms of values of the smallest eigenvalue of the reconstructed states, showing the (a) positive and (b) negative values.  Red circles indicate bins containing zero samples.  The grey, shaded area shows the region counted as zero for all the simulations carried out in this paper.}
\label{fig-counting-eigs}
\end{figure}

The number of eigenvalues in the reconstructed states can be counted by identifying a cut-off and counting as nonzero only those eigenvalues which are greater than that value.  Figure~\ref{fig-counting-eigs}(b-i) shows a histogram of the smallest eigenvalues for reconstructed two-qubit states from the same simulation described in Fig.~\ref{fig-penalty-function-mixture-trends}(d--e).  Histograms for all systems studied and for all eigenvalues (except the largest, which must by definition be larger than $1/d$) show a similar trend, namely some distribution of nonzero eigenvalues at larger values which tends to zero by around $10^{-6}$ and another large lobe of the distribution below a few $10^{-7}$ which can be identified as ``falsely'' nonzero due to some form of imprecision noise.  This is confirmed by the matching lobe in the histogram showing the elements of the distribution with negative values in Fig.~\ref{fig-counting-eigs}(b-ii).  This threshold, which is far above machine precision, may be a result of and set by tolerances in the reconstruction process itself, which is otherwise supposed to ensure that eigenvalues are all positive.

\begin{figure}
\begin{center}\begin{tabular}{ccc}
\includegraphics[width=55mm]{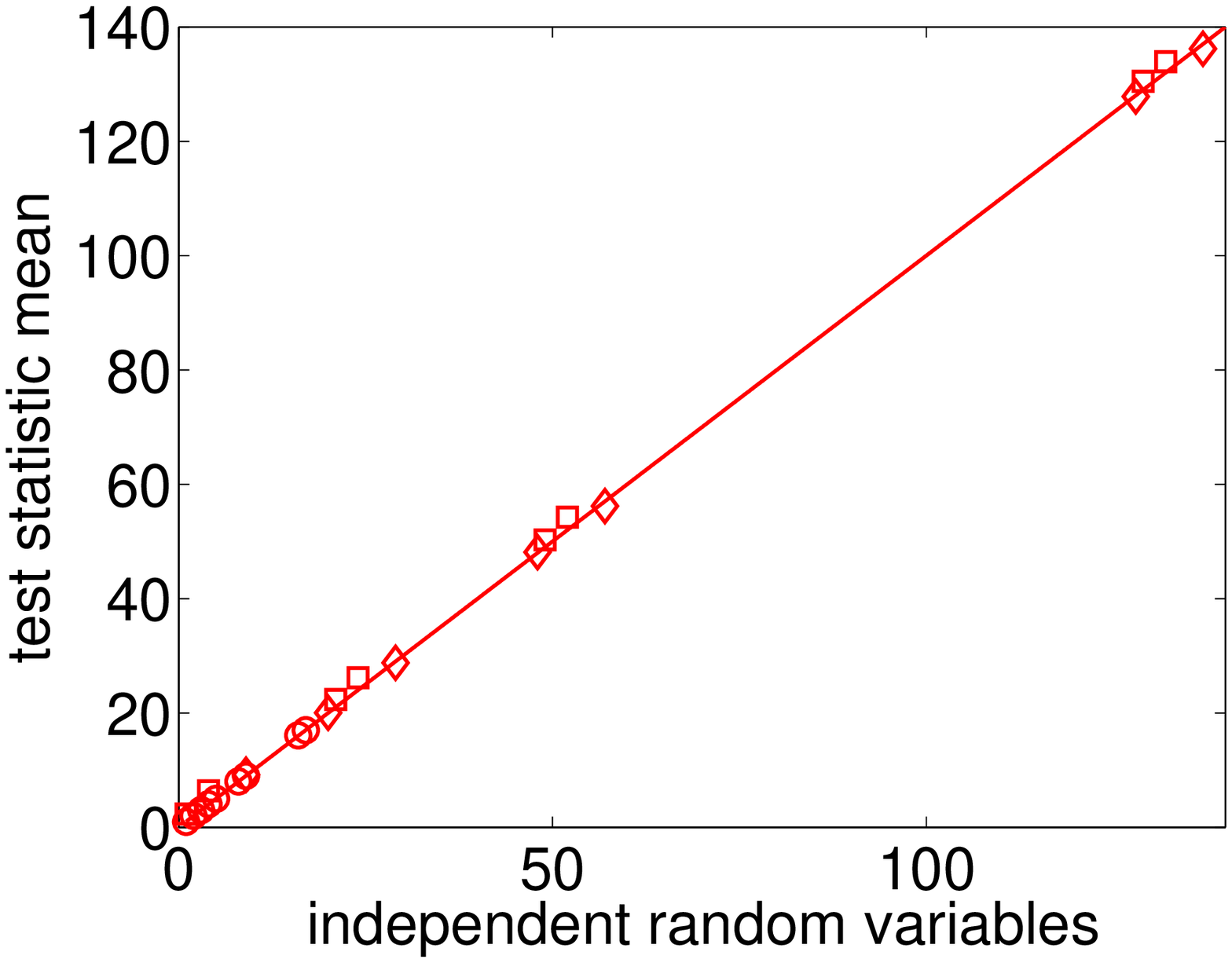} &\hspace{5mm}&
\includegraphics[width=55mm]{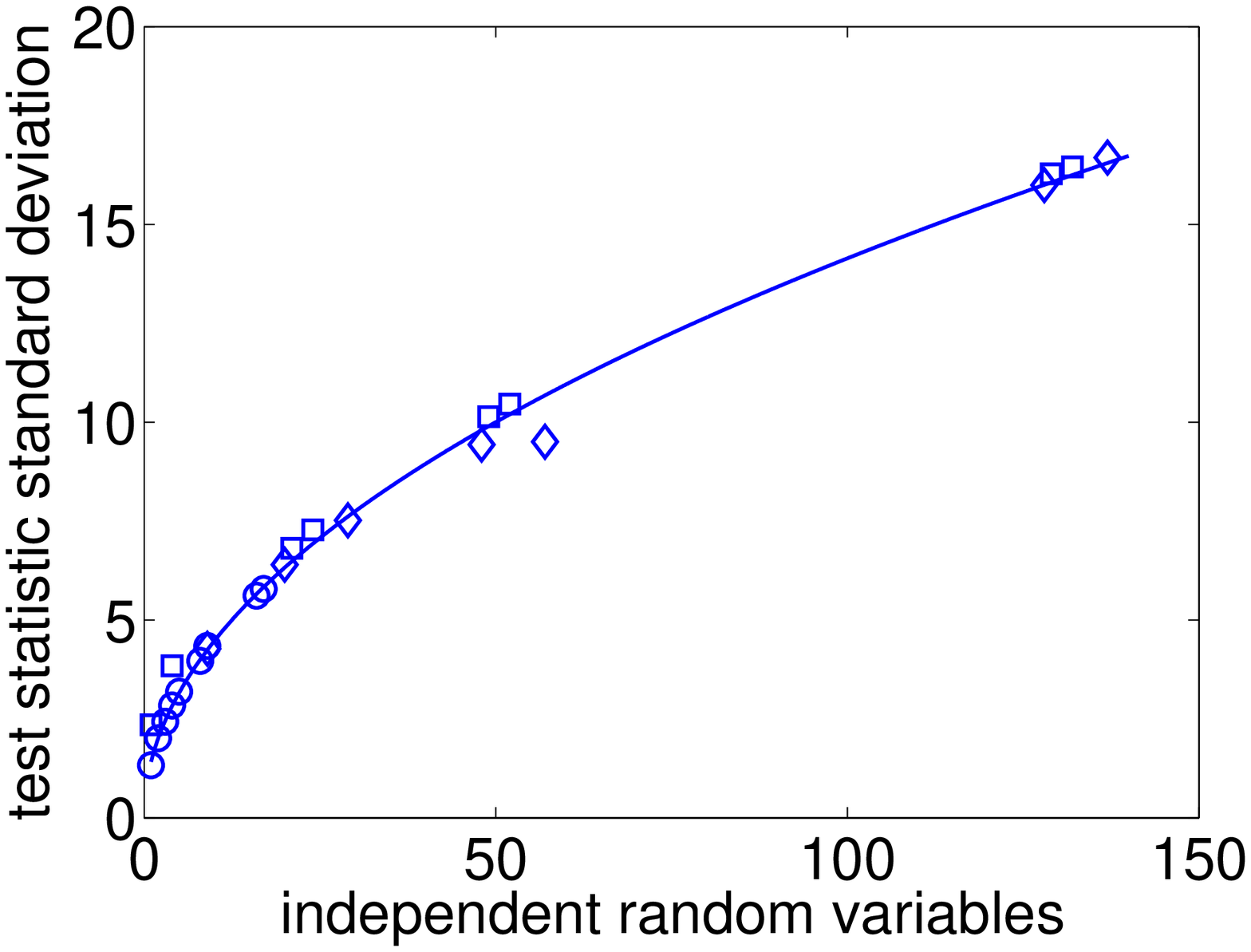} \\
\vspace{+2mm} (a) & \hspace{5mm} & (b)
\end{tabular} \\
\begin{tabular}{ccccccc}
\includegraphics[width=40mm]{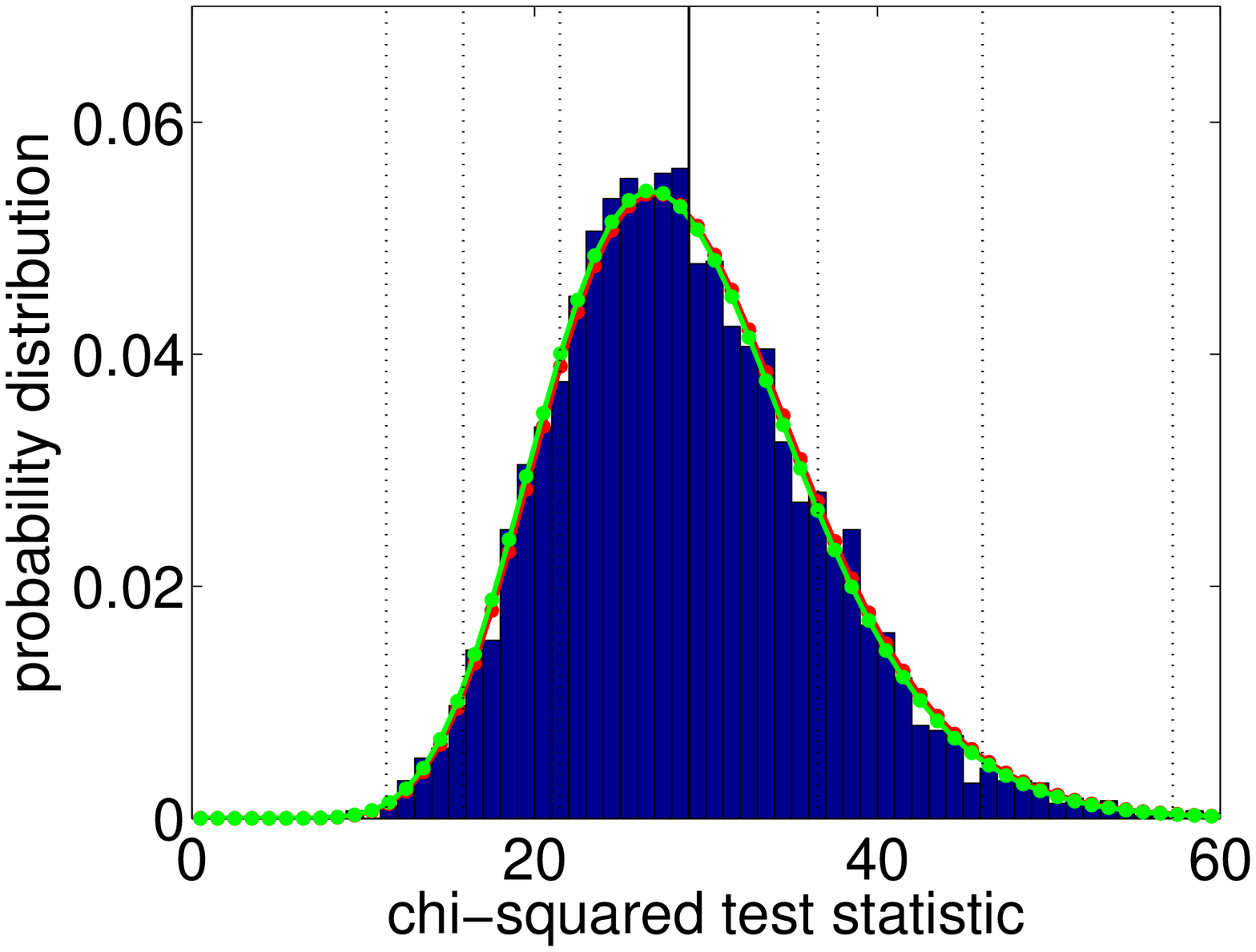} & \hspace{1mm} &
\includegraphics[width=40mm]{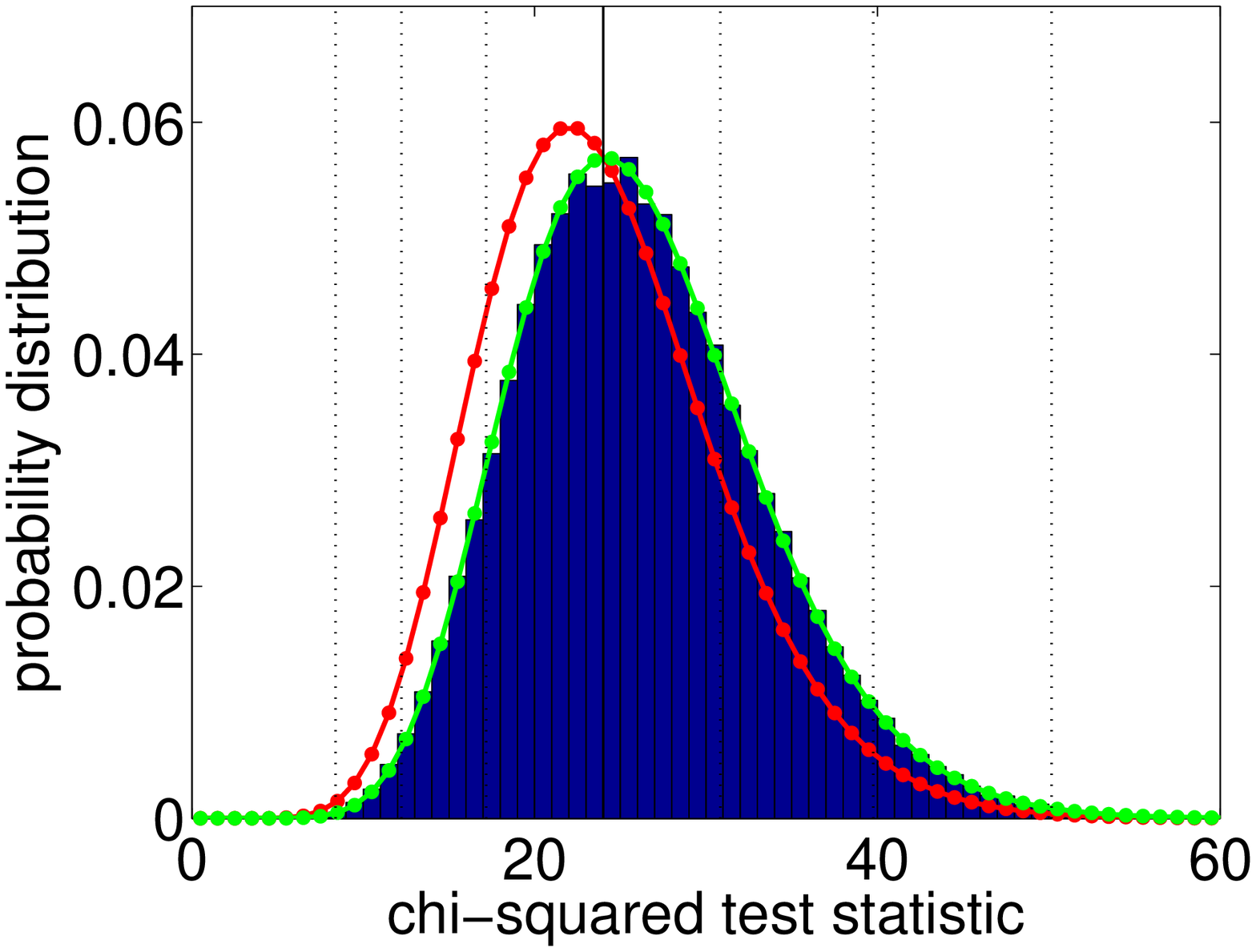} & \hspace{1mm} &
\includegraphics[width=40mm]{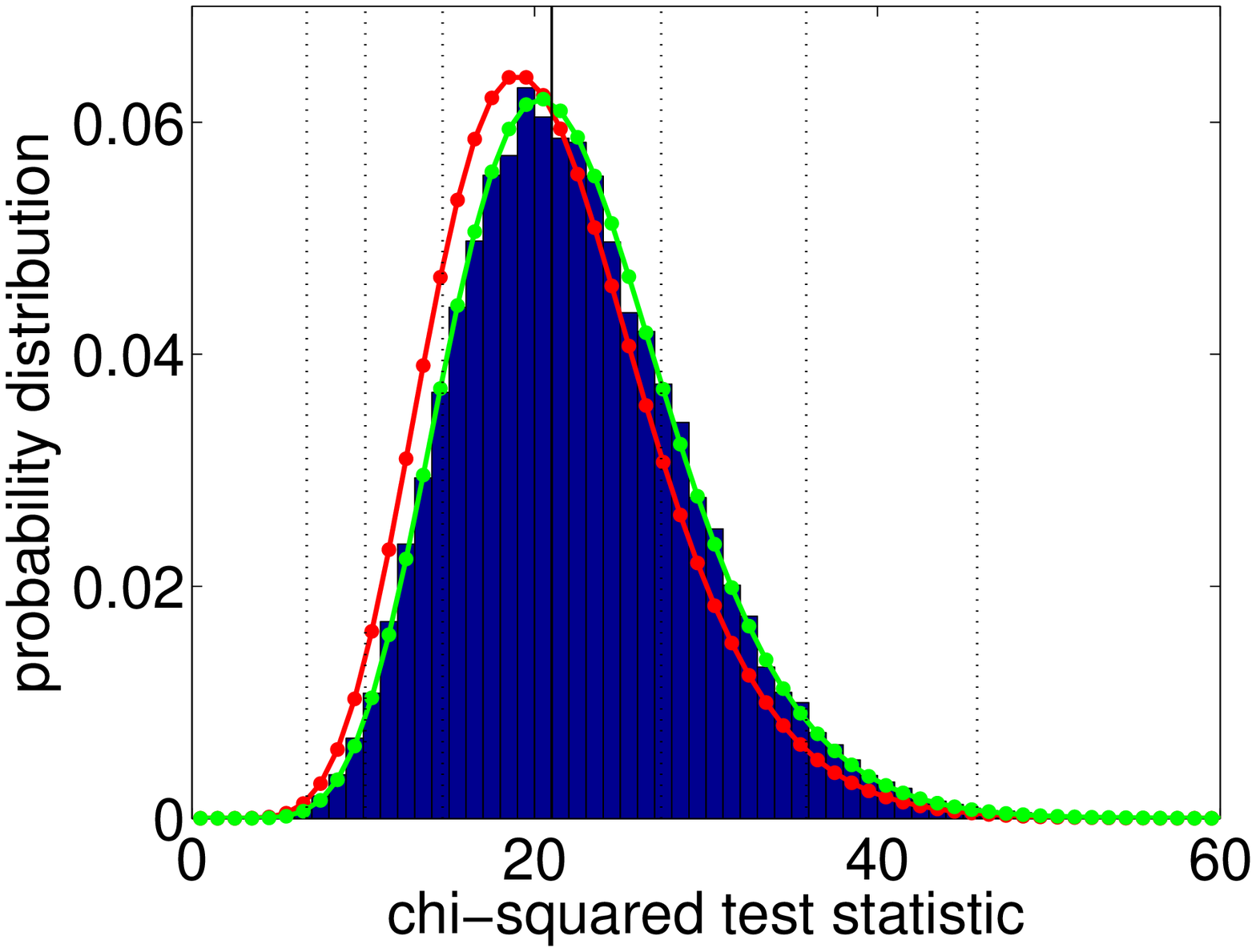} & \hspace{1mm} &
\includegraphics[width=40mm]{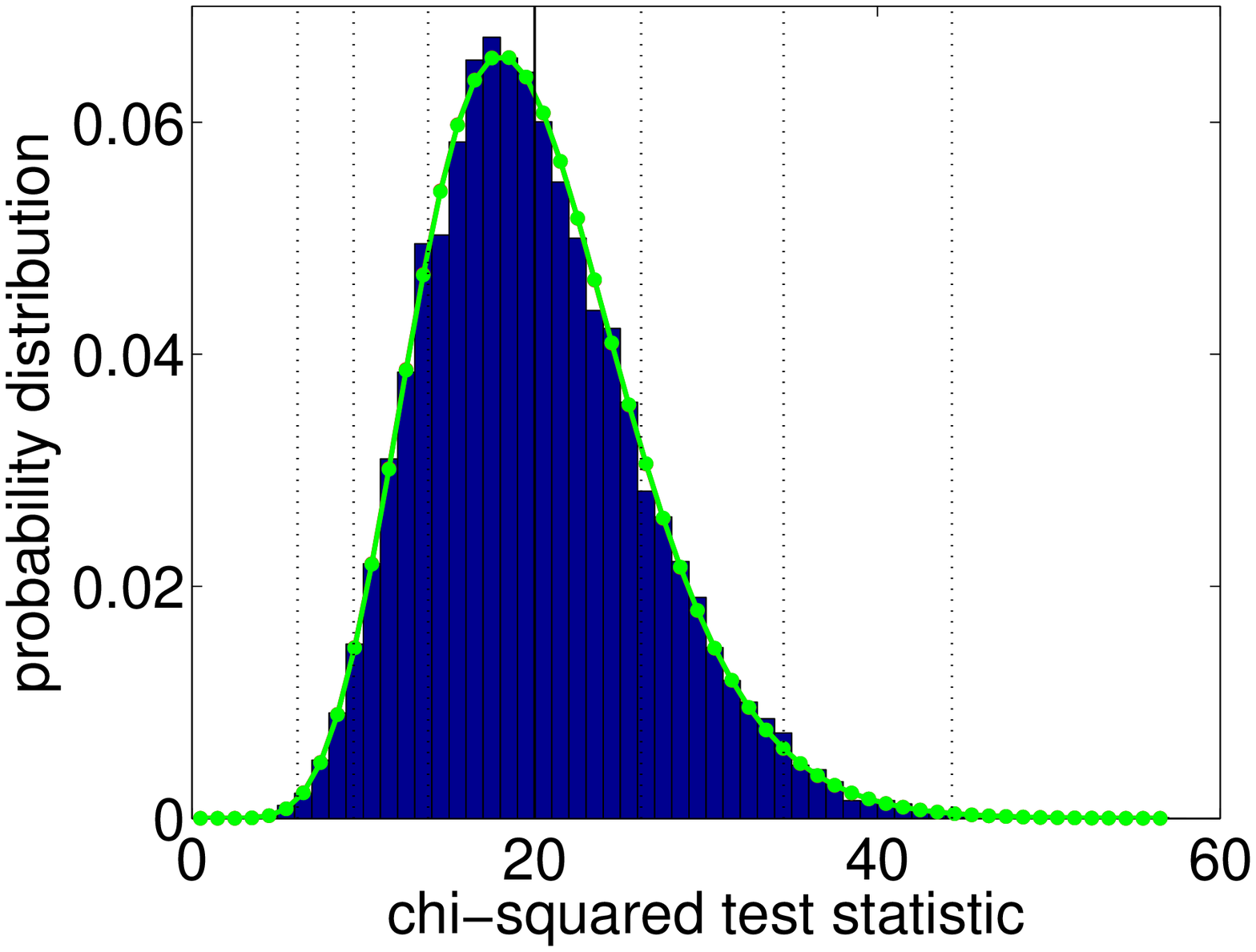} \\
\vspace{+2mm} (c-i) && (c-ii) && (c-iii) && (c-iv) \\
\end{tabular}\end{center}
\caption{Numerical simulations illustrating the dependence of the $X^2$ test statistic's distribution on the rank of the reconstructed density matrix.  Measured means (a) and standard deviations (b) of $X^2$ distributions for a range of systems (with different dimensions and measurement set sizes) with the reconstructed states grouped by the number of nonzero eigenvalues, plotted against the expected number of independent random variables, defined by Eq.~(\ref{eq-mixed-state-constraints}).  Solid curves show the expected values of $M{-}c$ and $\sqrt{M{-}c}$.  The circles ($\circ$) show single-qubit simulations with 6, 8, 12 and 20 measurement settings; the other points show two-qubit simulations with 6, 8 and 12 measurements per qubit ($M=36$, $M=64$ and $M=144$) for the pure-state and full-rank reconstructions (diamonds: $\Diamond$) and 2 and 3 nonzero eigenvalues (squares: $\Box$).  (c) Distribution of the test statistic for 150000 randomly chosen, mixed two-qubit state reconstructions with 6 measurements per qubit, grouped by the number of nonzero eigenvalues: (i) one, (ii) two, (iii) three and (iv) four (full-rank) nonzero eigenvalues.  Red curves show $\chi^2$ probability distributions for the expected number of independent parameters and green curves show the optimal $\chi^2$ distributions with a fit to the number of DOFs (over arbitrary real numbers).}
\label{fig-penalty-function-eigs}
\end{figure}

Figures~\ref{fig-penalty-function-eigs}(a) and (b) show the mean and standard deviation of the $X^2$ test statistic distributions obtained by grouping the reconstructed states by number of nonzero eigenvalues, plotted against the number of independent variables defined by Eq.~(\ref{eq-mixed-state-constraints}).  The results show generally good agreement with the theoretical curves which indicate the values expected for a $\chi^2$ distribution with the appropriate number of parameters, in particular for all those cases corresponding to either pure states or full-rank mixed states.  In the two-qubit tomography, however, the cases corresponding to 2 or 3 nonzero eigenvalues, while in the right region, do not provide such good agreement.  To illustrate this in more detail, Fig.~\ref{fig-penalty-function-eigs}(c) shows the observed distribution for the two-qubit simulations with six measurements per qubit ($M=36$).  There is excellent agreement of the optimal (green) and expected (red) theoretical distributions for Figs~\ref{fig-penalty-function-eigs}(c-i) and (c-iv), but a visible disparity in Figs~\ref{fig-penalty-function-eigs}(c-ii) and (c-iii).  This trend was relatively consistent for all systems studied.

These observed discrepancies are most pronounced when considering minimally complete tomographic reconstructions, such as those based on the tetrahedron single-qubit measurement set.  When the state is sufficiently far from the state-space boundary, there are enough free parameters in the resulting full-rank reconstruction to allow a perfect fit to the measured data, and the measured test statistic is always zero.  Near the boundary, however, when the reconstruction has fewer nonzero eigenvalues, the physicality constraints prevent a perfect fit to the data and the test statistic becomes again nonzero.  In this case, because the full-rank reconstructions have no DOFs, this in some sense isolates these discrepancies from competing factors and the change in behaviour is much more pronounced.  Studying this may therefore provide key insight into the general problem.   Nevertheless, evidence from other performance measures (e.g.,~\cite{deBurghMD2008}) strongly suggests that experimentalists should in any case avoid using minimally complete measurement sets for tomography.  Fortunately, in other cases, the observed discrepancies are generally substantially smaller than the width of the $X^2$ distributions and therefore do not greatly limit the usefulness of the chi-squared quality parameters for providing quantitative validation of experimental tomography results, as discussed in the next section.

\begin{figure}
\begin{center}\begin{tabular}{ccc}
(a) & \hspace{5mm} & (b) \\
\includegraphics[width=50mm]{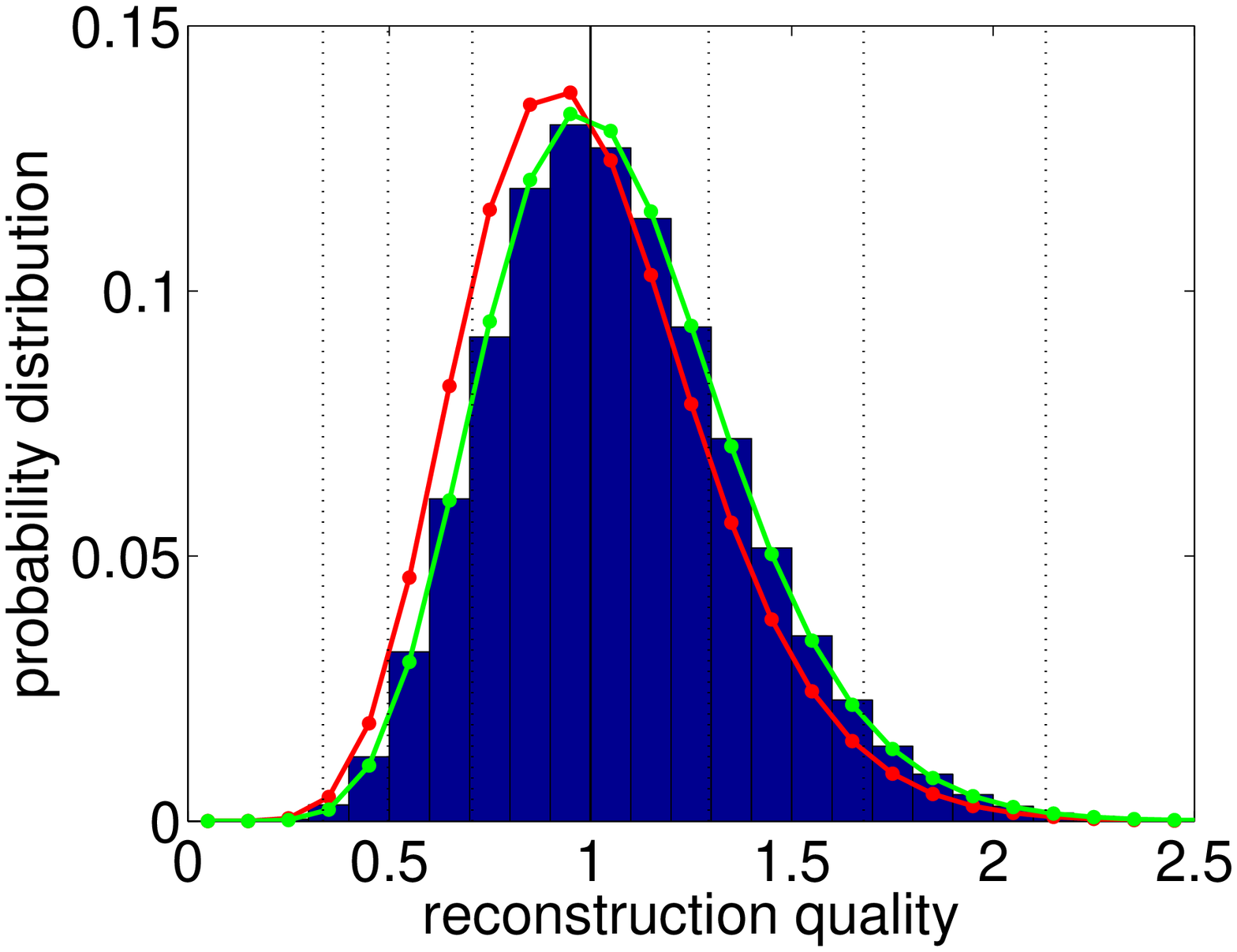} &&
\includegraphics[width=50mm]{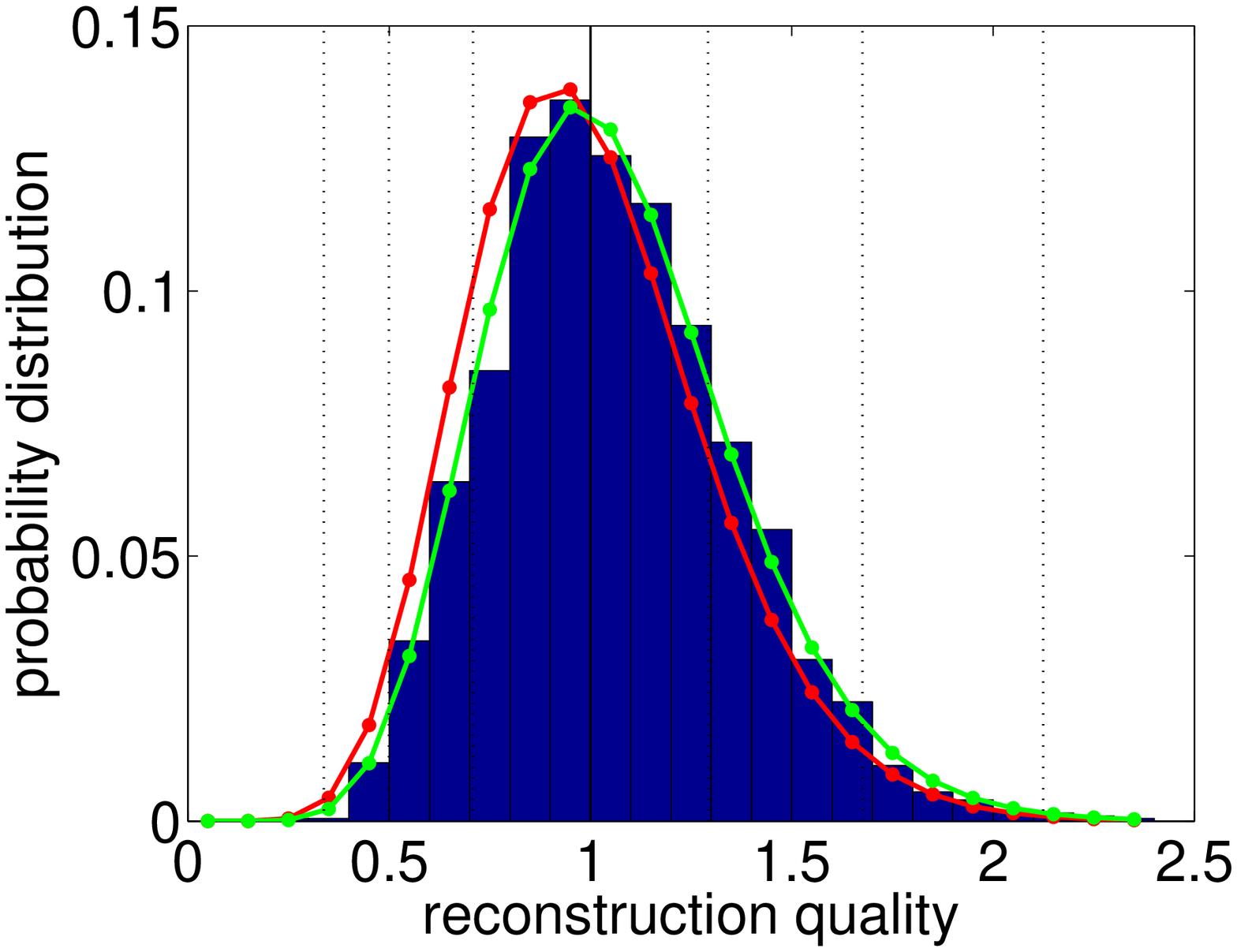}
\end{tabular}\end{center}
\caption{Reduced chi-squared quality parameter distributions with the number of constraints: (a) determined by counting the nonzero eigenvalues in the reconstructed states; and (b) calculated as an average estimated using Monte-Carlo simulations.}
\label{fig-fit-quality-dist-2QB}
\end{figure}

Finally, for a given set of experimental data, the reconstruction quality is calculated using Eq.~(\ref{eq-reduced-test-statistic}) with the number of constraints $c$ calculated by counting the number of nonzero eigenvalues in the reconstructed density matrix and applying Eq.~(\ref{eq-mixed-state-constraints}).  Figure~\ref{fig-fit-quality-dist-2QB}(a) shows the resulting distribution for the two-qubit, 36-measurement case discussed above.  Because of the observed discrepancy between the predicted and optimal distributions, $7.81\pm0.07\%$ ($1.79\pm0.03\%$) of simulated reconstructions had a reconstruction quality above the 95\% (99\%) confidence level threshold, instead of the 5\% (1\%) expected for data containing no excess noise.  This is nevertheless a substantial improvement over the na\"ive reconstruction quality calculated assuming $d^2$ constraints.

At this point, there is a related approach to analysing the reconstruction quality which is explicitly motivated by the assumption that the underlying statistics are governed by a general $\bar{\chi}^2$ distribution.  The $\bar{\chi}^2$ distribution is characterised by its so-called survival probability function
\begin{align}
\label{eq-chibarsq-survival}
P(\bar{\chi}^2 > X^2) = \sum_j w_j P(\chi_j^2 > X^2),
\end{align}
where $w_j$ are the weights which define the shape of the underlying $\bar{\chi}^2$ distribution.  Knowledge of $w_j$ would therefore immediately allow the calculation of a $p$-value which characterises the reconstruction quality.  However, since determining $w_j$ analytically is known to be difficult~\cite{DykstraR1991,StoelRD2006}, the standard approach is to estimate them using a standard Monte-Carlo technique~\cite{SilvapulleMJ1996, StoelRD2006}.  Specifically, either the raw data or the reconstructed state can be used to generate many samples of noisy data, given some particular noise model such as the Poissonian statistics assumed in these simulations.  Each resulting tomography then gives a density matrix and the number of nonzero eigenvalues can be counted as described above.  The distribution of these outcomes is then used to estimate the weights, $w_j$, and a $p$-value can be calculated from Eq.~(\ref{eq-chibarsq-survival}), with DOFs for the $\chi_j^2$ distributions determined from Eq.~(\ref{eq-mixed-state-constraints}).  From Eq.~(\ref{eq-chibarsq-survival}), it is easy to show that the mean of the underlying distribution is given by~\cite{DykstraR1991}
\begin{align}
\label{eq-chibarsq-mean}
\bar{\kappa} \equiv \expect{\bar{\chi}^2} = \sum_j w_j \expect{\chi_j^2} = \sum_j w_j \kappa_j.
\end{align}
It is therefore still useful as an aid to intuition to define a normalised reconstruction quality using $\bar{\kappa}$ in Eq.~(\ref{eq-reduced-test-statistic}).  Figure~\ref{fig-fit-quality-dist-2QB}(b) shows the resulting distribution for 1000 random two-qubit states (with $M=36$).  This method is obviously more computationally intensive, since it relies on repeated Monte-Carlo simulations to calculate the number of DOFs, but since this is often required already to determine error bars for derived physical quantities, this may not be a substantial further inconvenience.  The performance of these two techniques will be compared in the next section.

\subsection{Applications: diagnosing systematic errors in experimental tomography}

Ultimately, the main goal of calculating a goodness-of-fit statistic like the reconstruction quality is to be able to detect and diagnose whether the observed results really can be explained by the proposed noise model used to calculate the tomographic errors.  As discussed already, in real-world experiments, there may be many sources of errors other than those introduced by the statistics of finite counting.  For example, for nondeterministic quantum sources, such as photon sources, there may be systematic drift in the brightness of the source, either from the source interaction itself or from simple alignment drift, especially since tomography often requires long counting times.  Alternatively, there may be systematic errors in the measurement settings, as is quite common in ion-trapping and superconducting systems~\cite{MoroderT2012}, or unknown setting-related distortion in detector efficiencies, such as might occur in an integrated photonic detector which has polarisation-dependent sensitivity~\cite{GerritsT2012}.  Or there may even be some form of extra random statistical noise which does not originate from the expected effects of finite counting times, such as squeezed states exhibiting super-Poissonian counting statistics~\cite{BreitenbachG1997}, experiments operating in a regime where the Poissonian approximation breaks down, or statistical fluctuations in critical operating parameters like magnetic flux in superconducting circuits.  Furthermore, these extra sources of error can occur over a range of time scales, from much smaller than the single-measurement count time to longer than the overall tomography time and the effect on the reconstruction can vary greatly as a result.

\begin{figure}
\begin{center}\begin{tabular}{ccc}
\includegraphics[width=45mm]{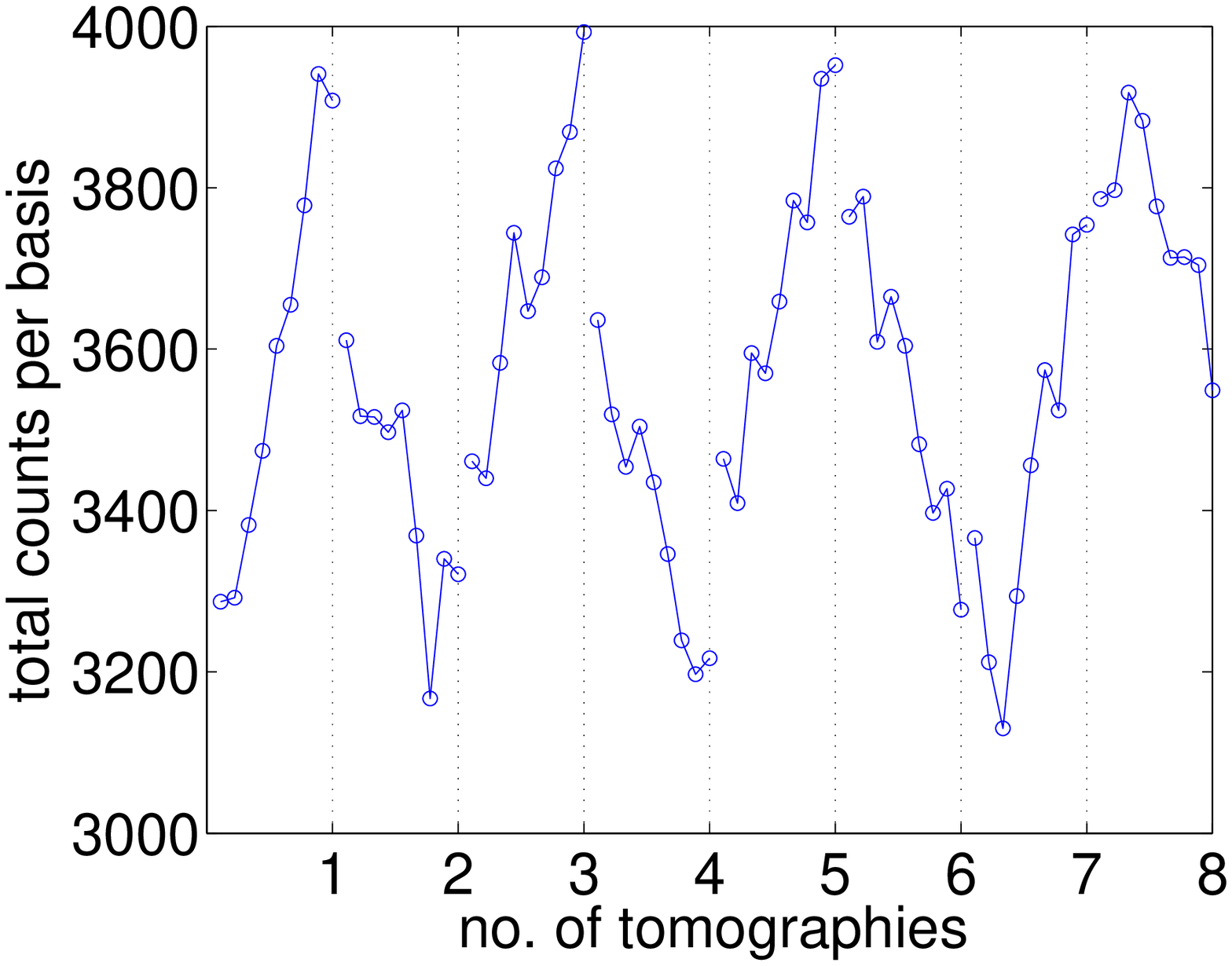} & \hspace{3mm} &
\includegraphics[width=45mm]{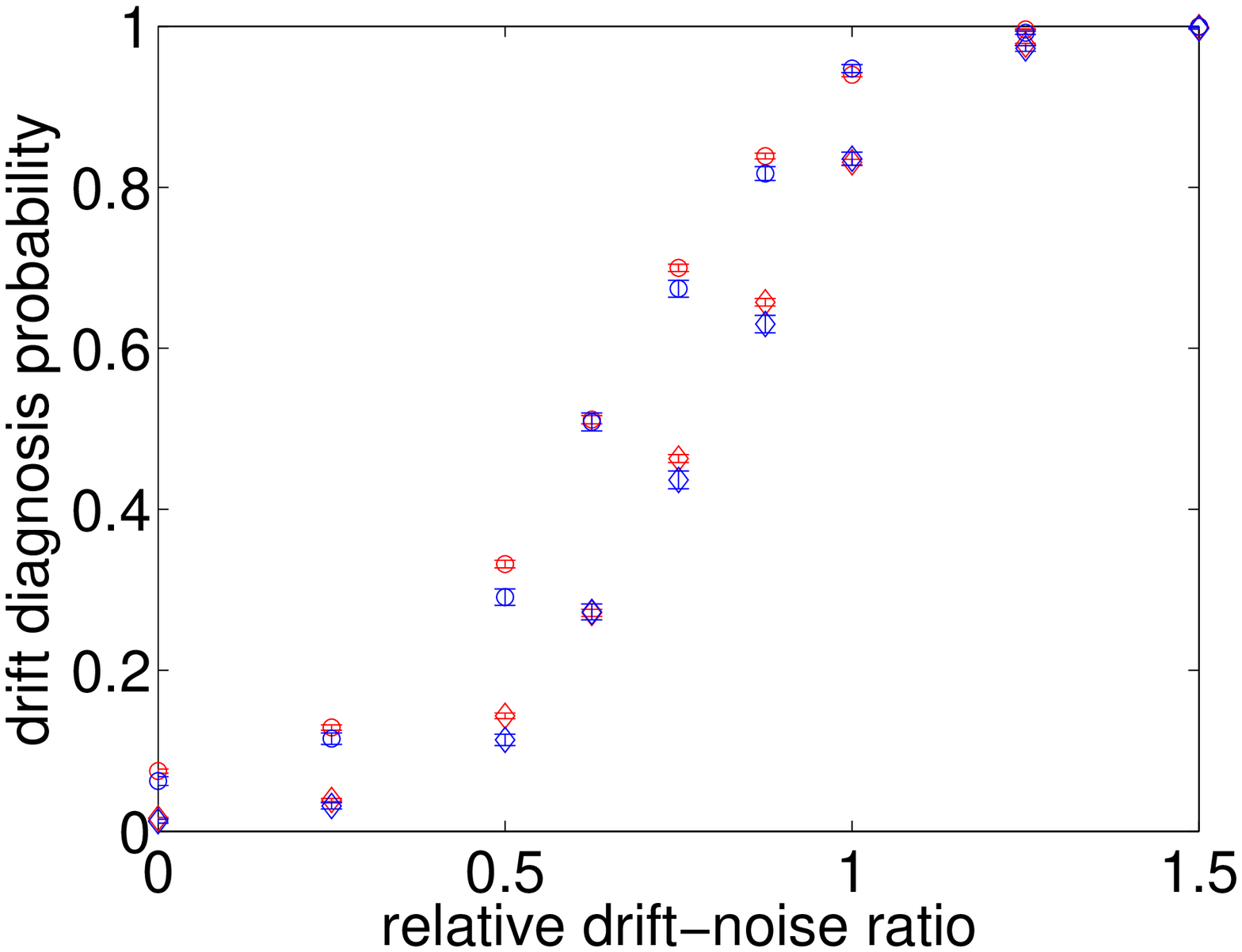}
\\
\vspace{+2mm} (a) && (b) \\
\end{tabular}\end{center}
\caption{(a) The effect of a laboratory air-conditioner on tomographic counts.  The counts show experimentally observed ``normalisations'' for a two-qubit state, determined by adding the total counts measured in four successive measurements of the orthogonal states in a given basis---there are 9 such sets in each two-qubit tomography made using 6 measurements per qubit.  The fluctuations are far larger than the expected Poissonian noise ($\sim$ 50 counts) and clear systematic cycling is apparent.  (b) The total probability of diagnosing the presence of additional noise versus the relative drift-noise ratio for 95\% ($\circ$) and 99\% ($\Box$) confidence using the reconstruction quality, $Q(\bar{\varrho})$, from Eq.~(\ref{eq-reduced-test-statistic}), with the number of constraints determined from Eq.~(\ref{eq-mixed-state-constraints}) either: (red) by simply counting the number of nonzero eigenvalues; or (blue) via Monte-Carlo simulation to calculate an average number for each set of data.}
\label{fig-fit-quality-drift}
\end{figure}

To illustrate the usefulness of the reconstruction quality, in this section, I study the first example of systematic drift in source brightness, one which is particularly problematic in photonic experiments.  Photonic experiments often use coupling to single-mode fibres or integrated chips to improve the visibility of quantum interference, but can become very sensitive to beam alignment as a result, even at the scale of temperature fluctuations induced by laboratory air-conditioning systems.  Often it is possible to ensure experimentally that the detailed conditions of the experiment are unchanged as a result (e.g., the source always produces the same state), but it is sometimes harder to eliminate drift in overall coupling efficiency and it is not uncommon to observe peak-to-peak fluctuations in brightness of 20\% or more [Fig.~\ref{fig-fit-quality-drift}(a)].

\begin{figure}
\begin{center}\begin{tabular}{ccccccc}
\includegraphics[width=40mm]{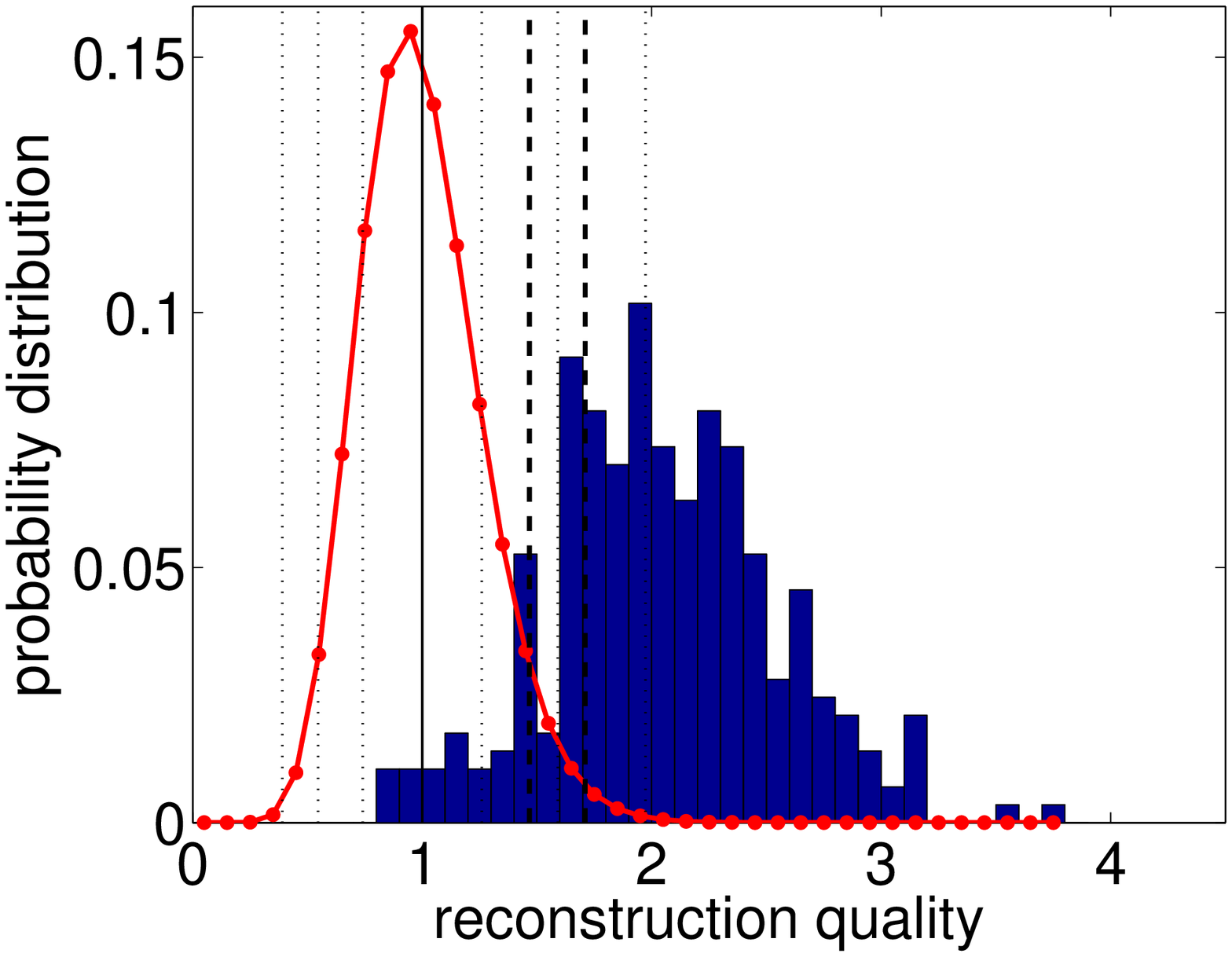} & \hspace{1mm} &
\includegraphics[width=40mm]{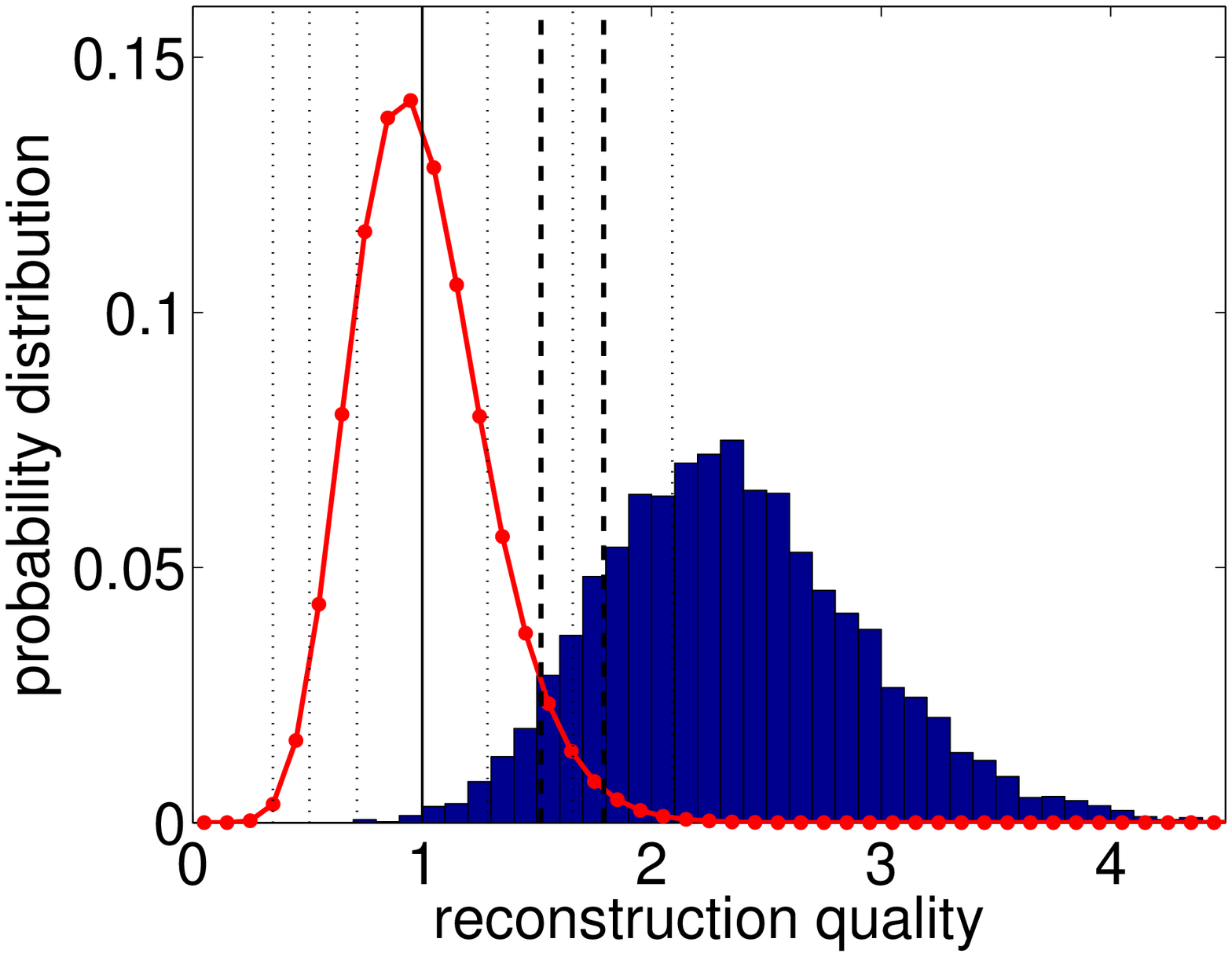} & \hspace{1mm} &
\includegraphics[width=40mm]{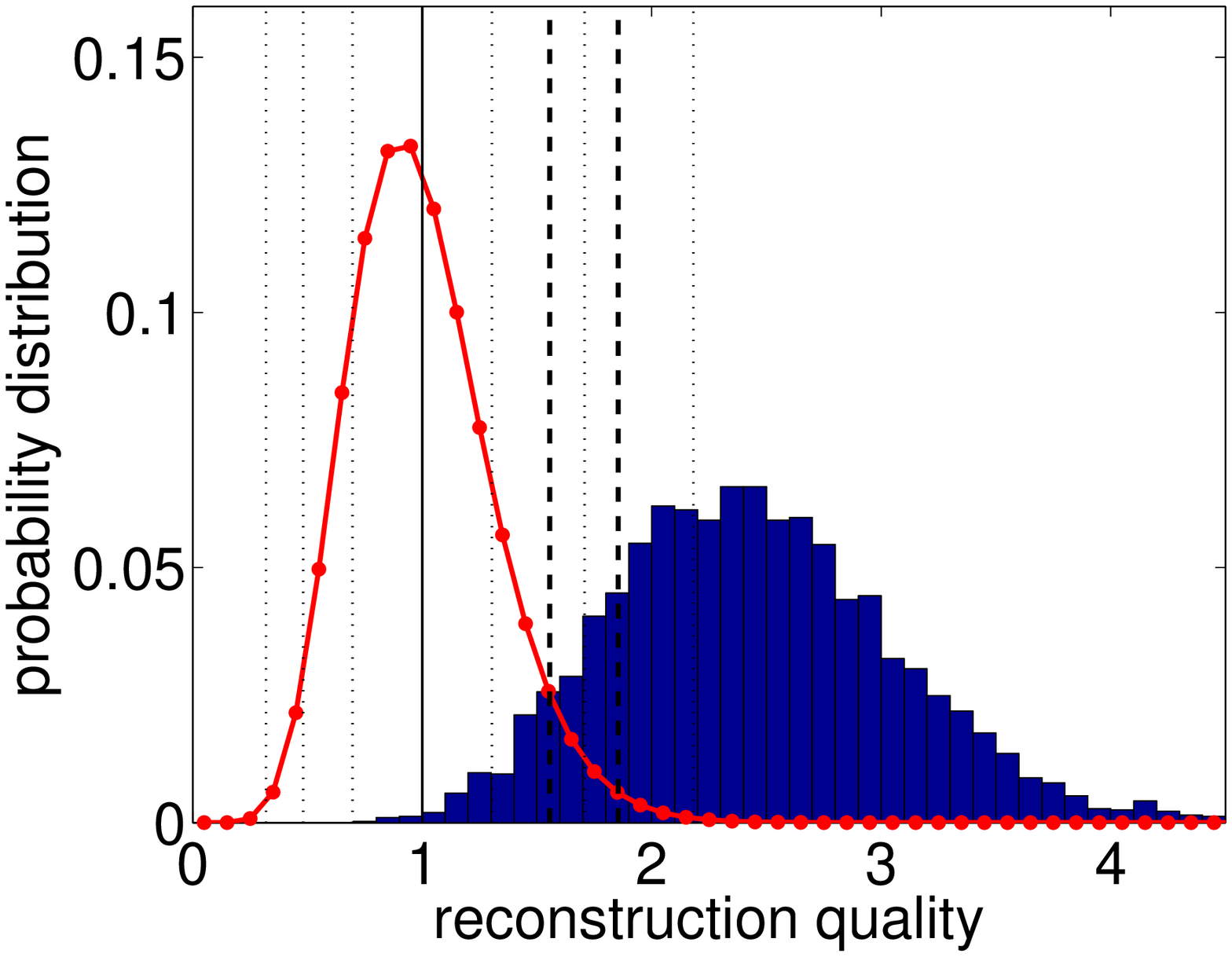} & \hspace{1mm} &
\includegraphics[width=40mm]{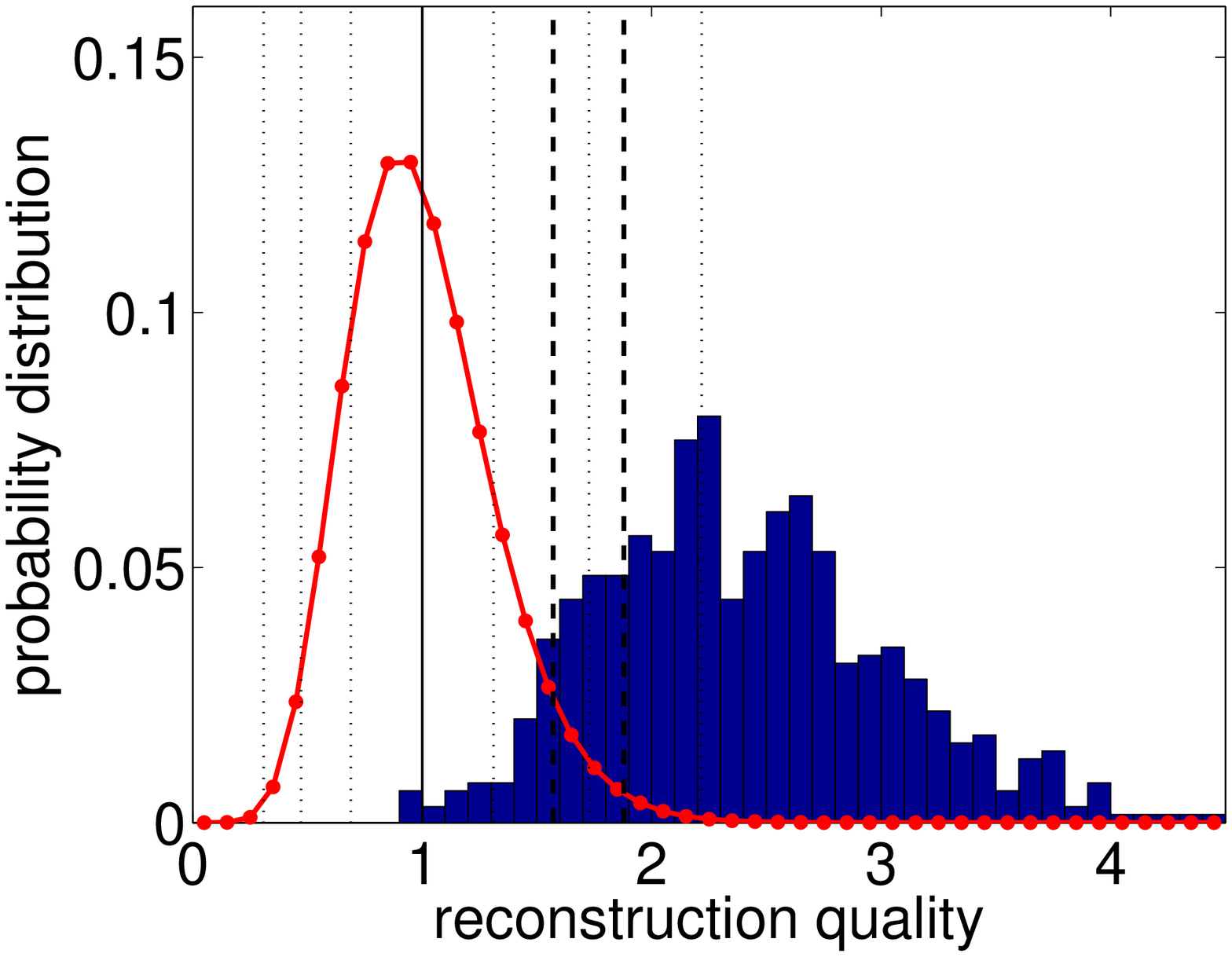} \\
\vspace{+2mm} (a) && (b) && (c) && (d) \\
\end{tabular}\end{center}
\caption{Distribution of the reconstruction quality for a source with systematic, sinusoidal fluctuations in source brightness with a relative drift-noise ratio of 1.0 and a period of 9.5 count settings.  Distributions calculated from 10,000 randomly chosen, mixed two-qubit state reconstructions with 6 measurements per qubit, grouped by the number of nonzero eigenvalues: (a) one, (b) two, (c) three and (d) four (full-rank) nonzero eigenvalues.  Red curves show $\chi^2$ probability distributions for the expected number of independent parameters.  Vertical lines show the expected mean (solid), the first few standard deviations on either side of the mean (dotted) and the cut-off values for diagnosing the presence of unexpected noise with 95\% and 99\% confidence.}
\label{fig-fit-quality-dist-drift-2QB}
\end{figure}

Using the familiar case of random photonic, two-qubit mixed states and 6 measurements per qubit, a simulated systematic, sinusoidal fluctuation in source brightness is added to the ``ideal'' counts before the simulated Poissonian noise is finally added on top.  The sensitivity of a tomography to such systematic fluctuations is determined by the amplitude of those fluctuations relative to the average size of the statistical noise.  Therefore, to make the scale meaningful, the relative amplitude of the systematic fluctuations is set in proportion to average expected standard deviation of the statistical noise ($1/\sqrt{N}$) over all counts in the simulation.  Figure~\ref{fig-fit-quality-dist-drift-2QB} shows the normalised quality distributions for a relative drift-noise ratio of 1.0 (with a period of 9.5 count settings) for 10,000 random states with a mean normalisation of 2000 overall photons per measurement setting, separated according to number of nonzero eigenvalues, and with the vertical lines showing the expected mean of 1 and the first few standard deviations for the corresponding reduced $\chi^2$ distribution.  Each case shows a shift which is clearly observable in the distributions of the quality parameter.  For each reconstruction, the reconstruction quality is then compared against the cut-off set by the desired $\chi^2$ confidence interval for the appropriate number of independent variables.  The total probability of diagnosing the presence of additional noise is plotted against the relative drift-noise ratio for two typical desired confidences of 95\% and 99\% in Fig.~\ref{fig-fit-quality-drift}(b) (red data).  Thus, with this type of drift, using the fit quality would diagnose the presence of unknown noise with 95\% confidence for more than 50\% of tomographic reconstructions given a system that exhibits brightness fluctuations which are only around 60\% the size of the statistical noise.  Note that the overall diagnosis rates will depend somewhat on the specific noise parameters, such as its oscillation period.  The diagnosis probabilities for zero noise are $0.075\pm0.003$ and $0.016\pm0.001$ for 95\% and 99\% confidence, respectively, indicative of a small, but observable effect from the discrepancies between expected and observed $X^2$ distributions for rank-deficient reconstructed states (discussed above).

A similar numerical study was made using the alternative Monte-Carlo technique for assessing the reconstruction quality.  Using the same system and conditions, for each of 1000 data sets, 100 samples were used to estimate the underlying $\bar{\chi}^2$ distribution and calculate the $p$-value which directly assessed the reconstruction quality.  These results are also plotted in Fig.~\ref{fig-fit-quality-drift}(b) (blue data) and show similar performance in general to the simpler technique.  The diagnosis probabilities for zero noise are $0.063\pm0.005$ and $0.013\pm0.002$ for 95\% and 99\% confidence, respectively, again slightly more than the expected values of 5\% and 1\%.

For both of these methods, the results show rates of over-diagnosis that are substantially smaller than those obtained based on making the assumption of $d^2$ constraints for all density matrices.  This demonstrates that making a more careful assessment of free parameters in the reconstruction greatly improves the faithfulness and usefulness of the standard and reduced chi-squared parameters as measures for diagnosing noise.

\subsection{Conclusions and future directions}

In this paper, I have outlined an intuitive approach to assessing reconstruction quality based on standard chi-squared statistical techniques.  I showed that the physicality constraints modify the number of free parameters available to the optimisation process, making it more complicated to predict the characteristics of the expected underlying $\bar{\chi}^2$ distribution, such as its mean.  Ignoring this and na\"ively using the standard value of $d^2$ leads to strongly over-estimated noise detection, which negates the value of such measures to provide meaningful, quantitative technical noise diagnosis.  To address this problem, I have taken a simple, heuristic approach to accounting for the effects of physicality constraints based on counting the number of free parameters in rank-deficient mixed states.  Specifically, I have suggested two related ways of calculating a normalised test statistic and demonstrated that they perform substantially more reliably in typical scenarios than the na\"ive approach, particularly in tomographic reconstructions based on over-complete measurement sets.  One is very easy to calculate, while the other, though more computationally intensive, may provide a more reliable representation of the underlying $\bar{\chi}^2$ statistics.  Numerical simulations showed that both measures are capable of diagnosing the presence of systematic fluctuations in source brightness, a typical problem in modern photonic experiments, even at noise levels lower than the ever-present statistical noise.  Given the substantial difference in computational complexity of these two techniques, it would be interesting in the future to carry out a detailed comparison of their performance.  It is also worth emphasising that the same approach should also offer benefits when applied to other forms of hypothesis tests, such as the generalised likelihood ratio test introduced in Ref.~\cite{MoroderT2012}, where the same tendency towards over-diagnosis of noise would be expected~\cite{ElBarmiH1999}.

I suggest that such a measure of reconstruction quality should always be reported with tomographic results, since it is as important to interpreting the data as any Monte-Carlo estimates of errors in physical quantities derived from the density matrix.  In fact, it could be argued that Monte-Carlo error estimation is meaningless without the supporting information provided by the reconstruction quality or an equivalent, to verify that the residual discrepancy between the observed data and the output state can indeed be plausibly explained by the noise model used to derive the error estimates.

In future work, further investigation is necessary to explore the origin of the discrepancy between the simulation and expected distributions for rank-deficient matrices with more than one eigenvalue, and to study in more detail the dependence of this discrepancy on factors such as mixture and precise form of the target or reconstructed state.  In particular, a promising path to pursue could be to explore this behaviour more fully in the context of minimally complete sets, where only the effects resulting from the physicality constraints imposed during reconstruction will be present.

It would also be interesting to see how the key idea of assessing the quality of a reconstruction in some way might be generalised to the context of region-based or Bayesian quantum estimation (e.g.,~\cite{Blume-KohoutR2012,ChristandlM2012,Blume-KohoutR2010a,AudenaertKMR2009}).  Finally, I would note that the good agreement between the expected theory and the simulations with mixed Werner-like states also show that the reconstruction quality may already be suitable for use with other forms of tomography which don't produce rank-deficient output matrices (e.g.,~\cite{Blume-KohoutR2010b}).

Ultimately, if quantum tomography is to fulfil its purpose of providing a rigorously quantitative way to characterise a quantum system, we need a comprehensive recipe for assessing the quality of tomographic reconstructions.  This is critical, both in order to authenticate the reconstruction and any physical quantities derived from it, and also to provide a reliable and accepted way to diagnose the presence of noise not accounted for in the expected noise model.  I have shown that a critical challenge in achieving this goal arises from the difficulty with identifying and quantifying the effects of physicality constraints in the reconstruction process.  Developing a recipe which is both deductive and reliable in situations where the physicality constraints come into play represents a major open problem for the field.  Moreover, if tomography is going to be genuinely useful as an everyday laboratory tool for trouble-shooting and benchmarking performance in quantum information processing experiments, it is also vital that the recipe provides a measure which is simple and fast to calculate.  This work aims to highlight the importance of these requirements and to outline a partial solution which can be implemented immediately by experimentalists.

\subsection{Acknowledgements}

A substantial part of the earlier stages of this work was carried out during my PhD at the University of Queensland.  I would like to gratefully acknowledge helpful conversations with Alexei Gilchrist, Andrew White, Andrew Doherty, Aephraim Steinberg, Bill Munro, Matthias Kleinmann, Otfried G\"uhne and Michael Sprague.  I would particularly like to thank Robin Blume-Kohout for suggestions and discussions at earlier stages of the work, and Marco Barbieri and Joshua Nunn for valuable discussions and detailed feedback on the manuscript.  This work was partly supported by an EC Marie Curie Intra-European Fellowship (PIEF-GA-2010-275103).


\begin{thebibliography}{44}
\expandafter\ifx\csname natexlab\endcsname\relax\def\natexlab#1{#1}\fi
\expandafter\ifx\csname bibnamefont\endcsname\relax
  \def\bibnamefont#1{#1}\fi
\expandafter\ifx\csname bibfnamefont\endcsname\relax
  \def\bibfnamefont#1{#1}\fi
\expandafter\ifx\csname citenamefont\endcsname\relax
  \def\citenamefont#1{#1}\fi
\expandafter\ifx\csname url\endcsname\relax
  \def\url#1{\texttt{#1}}\fi
\expandafter\ifx\csname urlprefix\endcsname\relax\def\urlprefix{URL }\fi
\providecommand{\bibinfo}[2]{#2}
\providecommand{\eprint}[2][]{\url{#2}}

\bibitem[{\citenamefont{Clauser et~al.}(1969)\citenamefont{Clauser, Horne,
  Shimony, and Holt}}]{ClauserJF1969}
\bibinfo{author}{\bibfnamefont{J.~F.} \bibnamefont{Clauser}},
  \bibinfo{author}{\bibfnamefont{M.~A.} \bibnamefont{Horne}},
  \bibinfo{author}{\bibfnamefont{A.}~\bibnamefont{Shimony}}, \bibnamefont{and}
  \bibinfo{author}{\bibfnamefont{R.~A.} \bibnamefont{Holt}},
  \bibinfo{journal}{Phys.\ Rev.\ Lett.} \textbf{\bibinfo{volume}{23}},
  \bibinfo{pages}{880} (\bibinfo{year}{1969}).

\bibitem[{\citenamefont{Wiseman et~al.}(2007)\citenamefont{Wiseman, Jones, and
  Doherty}}]{WisemanHM2007}
\bibinfo{author}{\bibfnamefont{H.~M.} \bibnamefont{Wiseman}},
  \bibinfo{author}{\bibfnamefont{S.~J.} \bibnamefont{Jones}}, \bibnamefont{and}
  \bibinfo{author}{\bibfnamefont{A.~C.} \bibnamefont{Doherty}},
  \bibinfo{journal}{Phys.\ Rev.\ Lett.} \textbf{\bibinfo{volume}{98}},
  \bibinfo{pages}{140402} (\bibinfo{year}{2007}).

\bibitem[{\citenamefont{Klyachko et~al.}(2008)\citenamefont{Klyachko, Can,
  Binicioglu, and Shumovsky}}]{KlyachkoAA2008}
\bibinfo{author}{\bibfnamefont{A.~A.} \bibnamefont{Klyachko}},
  \bibinfo{author}{\bibfnamefont{M.~A.} \bibnamefont{Can}},
  \bibinfo{author}{\bibfnamefont{S.}~\bibnamefont{Binicioglu}},
  \bibnamefont{and} \bibinfo{author}{\bibfnamefont{A.~S.}
  \bibnamefont{Shumovsky}}, \bibinfo{journal}{Phys.\ Rev.\ Lett.}
  \textbf{\bibinfo{volume}{101}}, \bibinfo{pages}{020403}
  (\bibinfo{year}{2008}).

\bibitem[{\citenamefont{Shor}(1994)}]{ShorPW1994}
\bibinfo{author}{\bibfnamefont{P.~W.} \bibnamefont{Shor}}, in
  \emph{\bibinfo{booktitle}{Proceedings of the 35th Annual Symposium on the
  Foundations of Computer Science}} (\bibinfo{year}{1994}), p.
  \bibinfo{pages}{124}.

\bibitem[{\citenamefont{White et~al.}(2007)\citenamefont{White, Gilchrist,
  Pryde, O'Brien, Bremner, and Langford}}]{WhiteAG2007}
\bibinfo{author}{\bibfnamefont{A.~G.} \bibnamefont{White}},
  \bibinfo{author}{\bibfnamefont{A.}~\bibnamefont{Gilchrist}},
  \bibinfo{author}{\bibfnamefont{G.~J.} \bibnamefont{Pryde}},
  \bibinfo{author}{\bibfnamefont{J.~L.} \bibnamefont{O'Brien}},
  \bibinfo{author}{\bibfnamefont{M.~J.} \bibnamefont{Bremner}},
  \bibnamefont{and} \bibinfo{author}{\bibfnamefont{N.~K.}
  \bibnamefont{Langford}}, \bibinfo{journal}{J.\ Opt.\ Soc.\ Am.\ B}
  \textbf{\bibinfo{volume}{24}}, \bibinfo{pages}{172} (\bibinfo{year}{2007}).

\bibitem[{\citenamefont{Chuang and Nielsen}(1997)}]{ChuangIL1997}
\bibinfo{author}{\bibfnamefont{I.~L.} \bibnamefont{Chuang}} \bibnamefont{and}
  \bibinfo{author}{\bibfnamefont{M.~A.} \bibnamefont{Nielsen}},
  \bibinfo{journal}{J.\ Mod.\ Opt.} \textbf{\bibinfo{volume}{44}},
  \bibinfo{pages}{2455} (\bibinfo{year}{1997}).

\bibitem[{\citenamefont{Smithey et~al.}(1993)\citenamefont{Smithey, Beck,
  Raymer, and Faridani}}]{SmitheyDT1993}
\bibinfo{author}{\bibfnamefont{D.~T.} \bibnamefont{Smithey}},
  \bibinfo{author}{\bibfnamefont{M.}~\bibnamefont{Beck}},
  \bibinfo{author}{\bibfnamefont{M.~G.} \bibnamefont{Raymer}},
  \bibnamefont{and} \bibinfo{author}{\bibfnamefont{A.}~\bibnamefont{Faridani}},
  \bibinfo{journal}{Phys.\ Rev.\ Lett.} \textbf{\bibinfo{volume}{70}},
  \bibinfo{pages}{1244} (\bibinfo{year}{1993}).

\bibitem[{\citenamefont{Blume-Kohout}(2010{\natexlab{a}})}]{Blume-KohoutR2010a}
\bibinfo{author}{\bibfnamefont{R.}~\bibnamefont{Blume-Kohout}},
  \bibinfo{journal}{New J.\ Phys.} \textbf{\bibinfo{volume}{12}},
  \bibinfo{pages}{043034} (\bibinfo{year}{2010}{\natexlab{a}}).

\bibitem[{\citenamefont{Audenaert and Scheel}(2009)}]{AudenaertKMR2009}
\bibinfo{author}{\bibfnamefont{K.~M.~R.} \bibnamefont{Audenaert}}
  \bibnamefont{and} \bibinfo{author}{\bibfnamefont{S.}~\bibnamefont{Scheel}},
  \bibinfo{journal}{New J.\ Phys.} \textbf{\bibinfo{volume}{11}},
  \bibinfo{pages}{023028} (\bibinfo{year}{2009}).

\bibitem[{\citenamefont{Hradil}(1997)}]{HradilZ1997}
\bibinfo{author}{\bibfnamefont{Z.}~\bibnamefont{Hradil}},
  \bibinfo{journal}{Phys.\ Rev.\ A} \textbf{\bibinfo{volume}{44}},
  \bibinfo{pages}{R1561} (\bibinfo{year}{1997}).

\bibitem[{\citenamefont{James et~al.}(2001)\citenamefont{James, Kwiat, Munro,
  and White}}]{JamesDFV2001}
\bibinfo{author}{\bibfnamefont{D.~F.~V.} \bibnamefont{James}},
  \bibinfo{author}{\bibfnamefont{P.~G.} \bibnamefont{Kwiat}},
  \bibinfo{author}{\bibfnamefont{W.~J.} \bibnamefont{Munro}}, \bibnamefont{and}
  \bibinfo{author}{\bibfnamefont{A.~G.} \bibnamefont{White}},
  \bibinfo{journal}{Phys.\ Rev.\ A} \textbf{\bibinfo{volume}{64}},
  \bibinfo{pages}{052312} (\bibinfo{year}{2001}).

\bibitem[{\citenamefont{Blume-Kohout}(2010{\natexlab{b}})}]{Blume-KohoutR2010b}
\bibinfo{author}{\bibfnamefont{R.}~\bibnamefont{Blume-Kohout}},
  \bibinfo{journal}{Phys.\ Rev.\ Lett.} \textbf{\bibinfo{volume}{105}},
  \bibinfo{pages}{200504} (\bibinfo{year}{2010}{\natexlab{b}}).

\bibitem[{\citenamefont{Christandl and Renner}(2012)}]{ChristandlM2012}
\bibinfo{author}{\bibfnamefont{M.}~\bibnamefont{Christandl}} \bibnamefont{and}
  \bibinfo{author}{\bibfnamefont{R.}~\bibnamefont{Renner}},
  \bibinfo{journal}{Phys.\ Rev.\ Lett.} \textbf{\bibinfo{volume}{109}},
  \bibinfo{pages}{120403} (\bibinfo{year}{2012}).

\bibitem[{\citenamefont{Blume-Kohout}(2012)}]{Blume-KohoutR2012}
\bibinfo{author}{\bibfnamefont{R.}~\bibnamefont{Blume-Kohout}},
  \bibinfo{journal}{arXiv.org} \textbf{\bibinfo{volume}{quant-ph}},
  \bibinfo{pages}{1202.5270} (\bibinfo{year}{2012}).

\bibitem[{\citenamefont{Langford}(2007)}]{LangfordNK2007phd}
\bibinfo{author}{\bibfnamefont{N.~K.} \bibnamefont{Langford}}, Ph.D. thesis,
  \bibinfo{school}{University of Queensland} (\bibinfo{year}{2007}).

\bibitem[{\citenamefont{Moroder et~al.}(2012)\citenamefont{Moroder, Kleinmann,
  Schindler, Monz, G{\"u}hne, and Blatt}}]{MoroderT2012}
\bibinfo{author}{\bibfnamefont{T.}~\bibnamefont{Moroder}},
  \bibinfo{author}{\bibfnamefont{M.}~\bibnamefont{Kleinmann}},
  \bibinfo{author}{\bibfnamefont{P.}~\bibnamefont{Schindler}},
  \bibinfo{author}{\bibfnamefont{T.}~\bibnamefont{Monz}},
  \bibinfo{author}{\bibfnamefont{O.}~\bibnamefont{G{\"u}hne}},
  \bibnamefont{and} \bibinfo{author}{\bibfnamefont{R.}~\bibnamefont{Blatt}},
  \bibinfo{journal}{arXiv.org} \textbf{\bibinfo{volume}{quant-ph}}
  (\bibinfo{year}{2012}).

\bibitem[{\citenamefont{Mood et~al.}(1974)\citenamefont{Mood, Graybill, and
  Boes}}]{Mood}
\bibinfo{author}{\bibfnamefont{A.~M.} \bibnamefont{Mood}},
  \bibinfo{author}{\bibfnamefont{F.~A.} \bibnamefont{Graybill}},
  \bibnamefont{and} \bibinfo{author}{\bibfnamefont{D.~C.} \bibnamefont{Boes}},
  \emph{\bibinfo{title}{Introduction to the Theory of Statistics}}
  (\bibinfo{publisher}{McGraw-Hill}, \bibinfo{address}{New York, NY, USA},
  \bibinfo{year}{1974}), \bibinfo{edition}{3rd} ed.

\bibitem[{\citenamefont{Taylor}(1997)}]{Taylor}
\bibinfo{author}{\bibfnamefont{J.~R.} \bibnamefont{Taylor}},
  \emph{\bibinfo{title}{An Introduction To Error Analysis}}
  (\bibinfo{publisher}{University Science Books}, \bibinfo{address}{Sausalito,
  CA}, \bibinfo{year}{1997}), \bibinfo{edition}{2nd} ed.

\bibitem[{\citenamefont{Blume-Kohout et~al.}(2010)\citenamefont{Blume-Kohout,
  Yin, and van Enk}}]{Blume-KohoutR2010c}
\bibinfo{author}{\bibfnamefont{R.}~\bibnamefont{Blume-Kohout}},
  \bibinfo{author}{\bibfnamefont{J.~O.~S.} \bibnamefont{Yin}},
  \bibnamefont{and} \bibinfo{author}{\bibfnamefont{S.~J.} \bibnamefont{van
  Enk}}, \bibinfo{journal}{Phys.\ Rev.\ Lett.} \textbf{\bibinfo{volume}{105}},
  \bibinfo{pages}{170501} (\bibinfo{year}{2010}).

\bibitem[{\citenamefont{Nielsen and Chuang}(2000)}]{NielsenChuang}
\bibinfo{author}{\bibfnamefont{M.~A.} \bibnamefont{Nielsen}} \bibnamefont{and}
  \bibinfo{author}{\bibfnamefont{I.~L.} \bibnamefont{Chuang}},
  \emph{\bibinfo{title}{Quantum Computation and Quantum Information}}
  (\bibinfo{publisher}{Cambridge University Press},
  \bibinfo{address}{Cambridge, UK}, \bibinfo{year}{2000}).

\bibitem[{\citenamefont{Penrose}(1955)}]{PenroseR1955}
\bibinfo{author}{\bibfnamefont{R.}~\bibnamefont{Penrose}},
  \bibinfo{journal}{P.\ Camb.\ Philos.\ Soc.} \textbf{\bibinfo{volume}{51}},
  \bibinfo{pages}{406} (\bibinfo{year}{1955}).

\bibitem[{\citenamefont{Boyd and Vandenberghe}(2004)}]{BoydVandenberghe}
\bibinfo{author}{\bibfnamefont{S.}~\bibnamefont{Boyd}} \bibnamefont{and}
  \bibinfo{author}{\bibfnamefont{L.}~\bibnamefont{Vandenberghe}},
  \emph{\bibinfo{title}{Convex Optimization}} (\bibinfo{publisher}{Cambridge
  University Press}, \bibinfo{address}{Cambridge, UK}, \bibinfo{year}{2004}).

\bibitem[{\citenamefont{Kosut et~al.}(2004)\citenamefont{Kosut, Walmsley, and
  Rabitz}}]{KosutRL2004}
\bibinfo{author}{\bibfnamefont{R.~L.} \bibnamefont{Kosut}},
  \bibinfo{author}{\bibfnamefont{I.~A.} \bibnamefont{Walmsley}},
  \bibnamefont{and} \bibinfo{author}{\bibfnamefont{H.}~\bibnamefont{Rabitz}},
  \bibinfo{journal}{arXiv.org} pp. \bibinfo{pages}{quant--ph/0411093}
  (\bibinfo{year}{2004}).

\bibitem[{\citenamefont{Kosut}(2008)}]{KosutRL2008}
\bibinfo{author}{\bibfnamefont{R.~L.} \bibnamefont{Kosut}},
  \bibinfo{journal}{arXiv.org} pp. \bibinfo{pages}{quant--ph/0812.4323}
  (\bibinfo{year}{2008}).

\bibitem[{\citenamefont{Gross et~al.}(2010)\citenamefont{Gross, Liu, Flammia,
  Becker, and Eisert}}]{GrossD2010}
\bibinfo{author}{\bibfnamefont{D.}~\bibnamefont{Gross}},
  \bibinfo{author}{\bibfnamefont{Y.-K.} \bibnamefont{Liu}},
  \bibinfo{author}{\bibfnamefont{S.}~\bibnamefont{Flammia}},
  \bibinfo{author}{\bibfnamefont{S.}~\bibnamefont{Becker}}, \bibnamefont{and}
  \bibinfo{author}{\bibfnamefont{J.}~\bibnamefont{Eisert}},
  \bibinfo{journal}{Phys.\ Rev.\ Lett.} \textbf{\bibinfo{volume}{105}},
  \bibinfo{pages}{150401} (\bibinfo{year}{2010}).

\bibitem[{\citenamefont{Flammia et~al.}(2012)\citenamefont{Flammia, Gross, Liu,
  and Eisert}}]{FlammiaST2012}
\bibinfo{author}{\bibfnamefont{S.~T.} \bibnamefont{Flammia}},
  \bibinfo{author}{\bibfnamefont{D.}~\bibnamefont{Gross}},
  \bibinfo{author}{\bibfnamefont{Y.-K.} \bibnamefont{Liu}}, \bibnamefont{and}
  \bibinfo{author}{\bibfnamefont{J.}~\bibnamefont{Eisert}},
  \bibinfo{journal}{arXiv.org} \textbf{\bibinfo{volume}{quant-ph}},
  \bibinfo{pages}{1205.2300} (\bibinfo{year}{2012}).

\bibitem[{\citenamefont{Shabani et~al.}(2011)\citenamefont{Shabani, Kosut,
  Mohseni, Rabitz, Broome, Almeida, Fedrizzi, and White}}]{ShabaniA2011}
\bibinfo{author}{\bibfnamefont{A.}~\bibnamefont{Shabani}},
  \bibinfo{author}{\bibfnamefont{R.~L.} \bibnamefont{Kosut}},
  \bibinfo{author}{\bibfnamefont{M.}~\bibnamefont{Mohseni}},
  \bibinfo{author}{\bibfnamefont{H.}~\bibnamefont{Rabitz}},
  \bibinfo{author}{\bibfnamefont{M.~A.} \bibnamefont{Broome}},
  \bibinfo{author}{\bibfnamefont{M.~P.} \bibnamefont{Almeida}},
  \bibinfo{author}{\bibfnamefont{A.}~\bibnamefont{Fedrizzi}}, \bibnamefont{and}
  \bibinfo{author}{\bibfnamefont{A.~G.} \bibnamefont{White}},
  \bibinfo{journal}{Phys.\ Rev.\ Lett.} \textbf{\bibinfo{volume}{106}},
  \bibinfo{pages}{100401} (\bibinfo{year}{2011}).

\bibitem[{\citenamefont{Doherty and Gilchrist}(2006)}]{DohertyAC2006}
\bibinfo{author}{\bibfnamefont{A.~C.} \bibnamefont{Doherty}} \bibnamefont{and}
  \bibinfo{author}{\bibfnamefont{A.}~\bibnamefont{Gilchrist}},
  \bibinfo{journal}{unpublished}  (\bibinfo{year}{2006}).

\bibitem[{Not()}]{Note-PoissonianCorrection}
\bibinfo{note}{This can be proved using the recursive differential technique
  described in the hint to Problem 11.9 in Ref.~\cite{Taylor}.}

\bibitem[{\citenamefont{Thomas-Peter et~al.}(2011)\citenamefont{Thomas-Peter,
  Langford, Datta, Zhang, Smith, Spring, Metcalf, Coldenstrodt-Ronge, Hu, Nunn
  et~al.}}]{ThomasPeterN2011}
\bibinfo{author}{\bibfnamefont{N.}~\bibnamefont{Thomas-Peter}},
  \bibinfo{author}{\bibfnamefont{N.~K.} \bibnamefont{Langford}},
  \bibinfo{author}{\bibfnamefont{A.}~\bibnamefont{Datta}},
  \bibinfo{author}{\bibfnamefont{L.}~\bibnamefont{Zhang}},
  \bibinfo{author}{\bibfnamefont{B.~J.} \bibnamefont{Smith}},
  \bibinfo{author}{\bibfnamefont{J.~B.} \bibnamefont{Spring}},
  \bibinfo{author}{\bibfnamefont{B.~J.} \bibnamefont{Metcalf}},
  \bibinfo{author}{\bibfnamefont{H.~B.} \bibnamefont{Coldenstrodt-Ronge}},
  \bibinfo{author}{\bibfnamefont{M.}~\bibnamefont{Hu}},
  \bibinfo{author}{\bibfnamefont{J.}~\bibnamefont{Nunn}}, \bibnamefont{et~al.},
  \bibinfo{journal}{New J.\ Phys.} \textbf{\bibinfo{volume}{13}},
  \bibinfo{pages}{055024} (\bibinfo{year}{2011}).

\bibitem[{\citenamefont{de~Burgh et~al.}(2008)\citenamefont{de~Burgh, Langford,
  Doherty, and Gilchrist}}]{deBurghMD2008}
\bibinfo{author}{\bibfnamefont{M.~D.} \bibnamefont{de~Burgh}},
  \bibinfo{author}{\bibfnamefont{N.~K.} \bibnamefont{Langford}},
  \bibinfo{author}{\bibfnamefont{A.~C.} \bibnamefont{Doherty}},
  \bibnamefont{and}
  \bibinfo{author}{\bibfnamefont{A.}~\bibnamefont{Gilchrist}},
  \bibinfo{journal}{Phys.\ Rev.\ A} \textbf{\bibinfo{volume}{78}},
  \bibinfo{pages}{052122} (\bibinfo{year}{2008}).

\bibitem[{\citenamefont{Langford et~al.}(2004)\citenamefont{Langford, Dalton,
  Harvey, O'Brien, Pryde, Gilchrist, Bartlett, and White}}]{LangfordNK2004}
\bibinfo{author}{\bibfnamefont{N.~K.} \bibnamefont{Langford}},
  \bibinfo{author}{\bibfnamefont{R.~B.} \bibnamefont{Dalton}},
  \bibinfo{author}{\bibfnamefont{M.~D.} \bibnamefont{Harvey}},
  \bibinfo{author}{\bibfnamefont{J.~L.} \bibnamefont{O'Brien}},
  \bibinfo{author}{\bibfnamefont{G.~J.} \bibnamefont{Pryde}},
  \bibinfo{author}{\bibfnamefont{A.}~\bibnamefont{Gilchrist}},
  \bibinfo{author}{\bibfnamefont{S.~D.} \bibnamefont{Bartlett}},
  \bibnamefont{and} \bibinfo{author}{\bibfnamefont{A.~G.} \bibnamefont{White}},
  \bibinfo{journal}{Phys.\ Rev.\ Lett.} \textbf{\bibinfo{volume}{93}},
  \bibinfo{pages}{053601} (\bibinfo{year}{2004}).

\bibitem[{\citenamefont{Werner}(1989)}]{WernerRF1989}
\bibinfo{author}{\bibfnamefont{R.~F.} \bibnamefont{Werner}},
  \bibinfo{journal}{Phys.\ Rev.\ A} \textbf{\bibinfo{volume}{40}},
  \bibinfo{pages}{4277} (\bibinfo{year}{1989}).

\bibitem[{\citenamefont{Shapiro}(1988)}]{ShapiroA1988}
\bibinfo{author}{\bibfnamefont{A.}~\bibnamefont{Shapiro}},
  \bibinfo{journal}{Int.\ Stat.\ Rev.} pp. \bibinfo{pages}{49--62}
  (\bibinfo{year}{1988}).

\bibitem[{\citenamefont{Barmi and Dykstra}(1999)}]{ElBarmiH1999}
\bibinfo{author}{\bibfnamefont{H.~E.} \bibnamefont{Barmi}} \bibnamefont{and}
  \bibinfo{author}{\bibfnamefont{R.~L.} \bibnamefont{Dykstra}},
  \bibinfo{journal}{J.\ Nonparametr.\ Stat.} \textbf{\bibinfo{volume}{11}},
  \bibinfo{pages}{233} (\bibinfo{year}{1999}).

\bibitem[{\citenamefont{Silvapulle}(1996)}]{SilvapulleMJ1996}
\bibinfo{author}{\bibfnamefont{M.~J.} \bibnamefont{Silvapulle}},
  \bibinfo{journal}{Stat.\ Probabil.\ Lett.} \textbf{\bibinfo{volume}{28}},
  \bibinfo{pages}{137} (\bibinfo{year}{1996}).

\bibitem[{\citenamefont{Barmi and Dykstra}(1995)}]{ElBarmiH1995}
\bibinfo{author}{\bibfnamefont{H.~E.} \bibnamefont{Barmi}} \bibnamefont{and}
  \bibinfo{author}{\bibfnamefont{R.}~\bibnamefont{Dykstra}},
  \bibinfo{journal}{Can.\ J.\ Stat.} \textbf{\bibinfo{volume}{23}},
  \bibinfo{pages}{131} (\bibinfo{year}{1995}).

\bibitem[{\citenamefont{Ling et~al.}(2006)\citenamefont{Ling, Soh,
  Lamas-Linares, and Kurtsiefer}}]{LingA2006}
\bibinfo{author}{\bibfnamefont{A.}~\bibnamefont{Ling}},
  \bibinfo{author}{\bibfnamefont{K.~P.} \bibnamefont{Soh}},
  \bibinfo{author}{\bibfnamefont{A.}~\bibnamefont{Lamas-Linares}},
  \bibnamefont{and}
  \bibinfo{author}{\bibfnamefont{C.}~\bibnamefont{Kurtsiefer}},
  \bibinfo{journal}{Phys.\ Rev.\ A} \textbf{\bibinfo{volume}{74}},
  \bibinfo{pages}{022309} (\bibinfo{year}{2006}).

\bibitem[{\citenamefont{Bengtsson and
  \.{Z}yczkowski}(2006)}]{BengtssonZyczkowski}
\bibinfo{author}{\bibfnamefont{I.}~\bibnamefont{Bengtsson}} \bibnamefont{and}
  \bibinfo{author}{\bibfnamefont{K.}~\bibnamefont{\.{Z}yczkowski}},
  \emph{\bibinfo{title}{Geometry of Quantum States: An Introduction to Quantum
  Entanglement}} (\bibinfo{publisher}{Cambridge University Press},
  \bibinfo{address}{Cambridge, UK}, \bibinfo{year}{2006}).

\bibitem[{\citenamefont{Adamson and Steinberg}(2010)}]{AdamsonRBA2010}
\bibinfo{author}{\bibfnamefont{R.~B.~A.} \bibnamefont{Adamson}}
  \bibnamefont{and} \bibinfo{author}{\bibfnamefont{A.~M.}
  \bibnamefont{Steinberg}}, \bibinfo{journal}{Phys.\ Rev.\ Lett.}
  \textbf{\bibinfo{volume}{105}}, \bibinfo{pages}{030406}
  (\bibinfo{year}{2010}).

\bibitem[{\citenamefont{Dykstra}(1991)}]{DykstraR1991}
\bibinfo{author}{\bibfnamefont{R.}~\bibnamefont{Dykstra}},
  \bibinfo{journal}{Can.\ J.\ Stat.} \textbf{\bibinfo{volume}{19}},
  \bibinfo{pages}{297} (\bibinfo{year}{1991}).

\bibitem[{\citenamefont{Stoel et~al.}(2006)\citenamefont{Stoel, Garre, Dolan,
  and van~den Wittenboer}}]{StoelRD2006}
\bibinfo{author}{\bibfnamefont{R.~D.} \bibnamefont{Stoel}},
  \bibinfo{author}{\bibfnamefont{F.~G.} \bibnamefont{Garre}},
  \bibinfo{author}{\bibfnamefont{C.}~\bibnamefont{Dolan}}, \bibnamefont{and}
  \bibinfo{author}{\bibfnamefont{G.}~\bibnamefont{van~den Wittenboer}},
  \bibinfo{journal}{Psychol.\ Methods} \textbf{\bibinfo{volume}{11}},
  \bibinfo{pages}{439} (\bibinfo{year}{2006}).

\bibitem[{\citenamefont{Gerrits et~al.}(2011)\citenamefont{Gerrits,
  Thomas-Peter, Gates, Lita, Metcalf, Calkins, Tomlin, Fox, Linares, Spring
  et~al.}}]{GerritsT2012}
\bibinfo{author}{\bibfnamefont{T.}~\bibnamefont{Gerrits}},
  \bibinfo{author}{\bibfnamefont{N.}~\bibnamefont{Thomas-Peter}},
  \bibinfo{author}{\bibfnamefont{J.~C.} \bibnamefont{Gates}},
  \bibinfo{author}{\bibfnamefont{A.~E.} \bibnamefont{Lita}},
  \bibinfo{author}{\bibfnamefont{B.~J.} \bibnamefont{Metcalf}},
  \bibinfo{author}{\bibfnamefont{B.}~\bibnamefont{Calkins}},
  \bibinfo{author}{\bibfnamefont{N.~A.} \bibnamefont{Tomlin}},
  \bibinfo{author}{\bibfnamefont{A.~E.} \bibnamefont{Fox}},
  \bibinfo{author}{\bibfnamefont{A.~L.} \bibnamefont{Linares}},
  \bibinfo{author}{\bibfnamefont{J.~B.} \bibnamefont{Spring}},
  \bibnamefont{et~al.}, \bibinfo{journal}{Phys.\ Rev.\ A}
  \textbf{\bibinfo{volume}{84}}, \bibinfo{pages}{060301(R)}
  (\bibinfo{year}{2011}).

\bibitem[{\citenamefont{Breitenbach et~al.}(1997)\citenamefont{Breitenbach,
  Schiller, and Mlynek}}]{BreitenbachG1997}
\bibinfo{author}{\bibfnamefont{G.}~\bibnamefont{Breitenbach}},
  \bibinfo{author}{\bibfnamefont{S.}~\bibnamefont{Schiller}}, \bibnamefont{and}
  \bibinfo{author}{\bibfnamefont{J.}~\bibnamefont{Mlynek}},
  \bibinfo{journal}{Nature} \textbf{\bibinfo{volume}{387}},
  \bibinfo{pages}{471} (\bibinfo{year}{1997}).

\end{thebibliography}

\end{document}